\documentclass[12pt] {article}
\usepackage{epsfig}
\def\aprle{\buildrel < \over {_{\sim}}} 
\def\aprge{\buildrel > \over {_{\sim}}} 
\def\lmag{\lambda_{{\rm mag}} }
\def\nubar{\overline{\nu}} 
\def\longa {{\overline{\varphi} }}

\begin{document}     

\title{The  fluxes of sub--cutoff particles  detected  by AMS, \\
       the cosmic ray albedo and  atmospheric  neutrinos}

\author{ Paolo Lipari  \\
\small   Dipartimento di Fisica, Universit\`a di Roma ``la Sapienza",\\
\small   and I.N.F.N., Sezione di Roma, P. A. Moro 2, I-00185 Roma, Italy
~~ \\
\small   also at: Research Center for Cosmic  Neutrinos,
  ICRR, University of Tokyo}

\date{\small January 31, 2001}          

\maketitle

\begin{abstract}
New  measurements  of the cosmic  ray  fluxes
 ($p$, $e^\pm$ and Helium)
performed by the Alpha Magnetic Spectrometer (AMS)  during
a ten days flight of  space shuttle 
 have revealed the existence of significant fluxes
of particles  below the  geomagnetic  cutoff. These   fluxes
exhibit a number of  remarkable properties,  such  as a
$^3$He/$^4$He ratio of order   $\sim 10$, an $e^+/e^-$
 ratio  of order   $\sim 4$ and
production from  well defined  regions of the Earth that are distinct
for  positively and  negatively charged particles.
In this work we show  that the natural  hypothesis,
that these subcutoff particles  are  generated as  secondary products of
primary cosmic  ray interactions  in the atmosphere  can reproduce
all the  observed properties.
We also  discuss the implications of the subcutoff fluxes 
for the estimate of the atmospheric   neutrino fluxes, and find that 
they represent  a negligibly small correction.
On the other hand the AMS   results  give  important
 confirmations about the assumption of 
isotropy for  the interplanetary   cosmic  ray  fluxes
also on   large angular scales,   
and on the validity of the geomagnetic  effects
that are important  elements for  the prediction of the atmospheric 
neutrino  fluxes.
\end{abstract}

\section{Introduction}
\begin{figure} [t]
\centerline{\psfig{figure=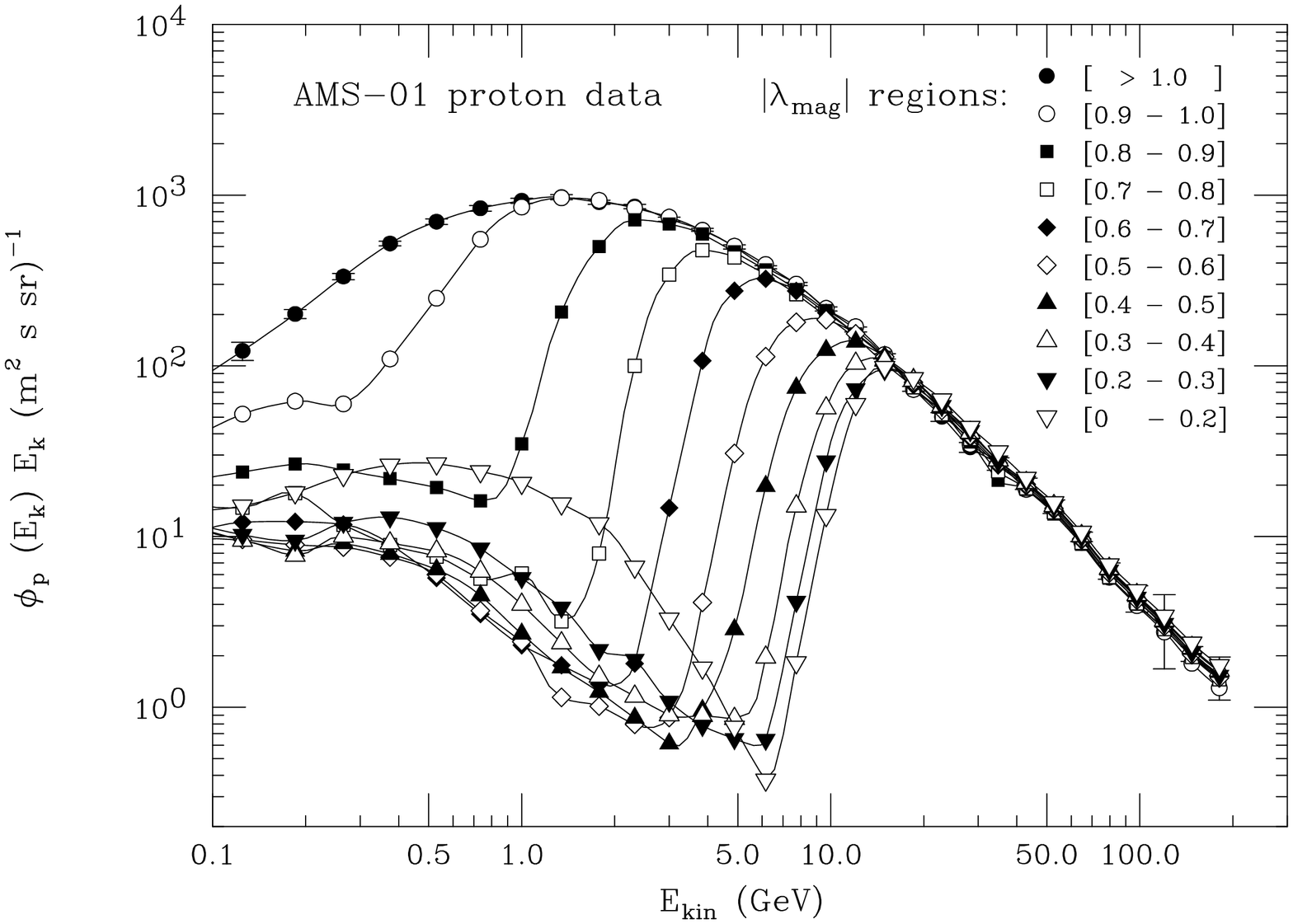,angle=90,height=12.9cm}}
\caption {\small  
Cosmic ray proton flux  measurements of the AMS  detector
 \protect\cite{AMS-protons}  averaged  for  a detector position
in ten regions  of  magnetic  longitude.
The error  bars are shown  only for one  set of measures
($|\lmag| > 1$~radiant). Note that
for $E_k > 0.3$~GeV  the sub--cutoff proton flux
is  most intense in the magnetic  equatorial region ($|\lmag| < 0.2$).
\label{fig:ams1}  }
\end{figure}

\begin{figure} [t]
\centerline{\psfig{figure=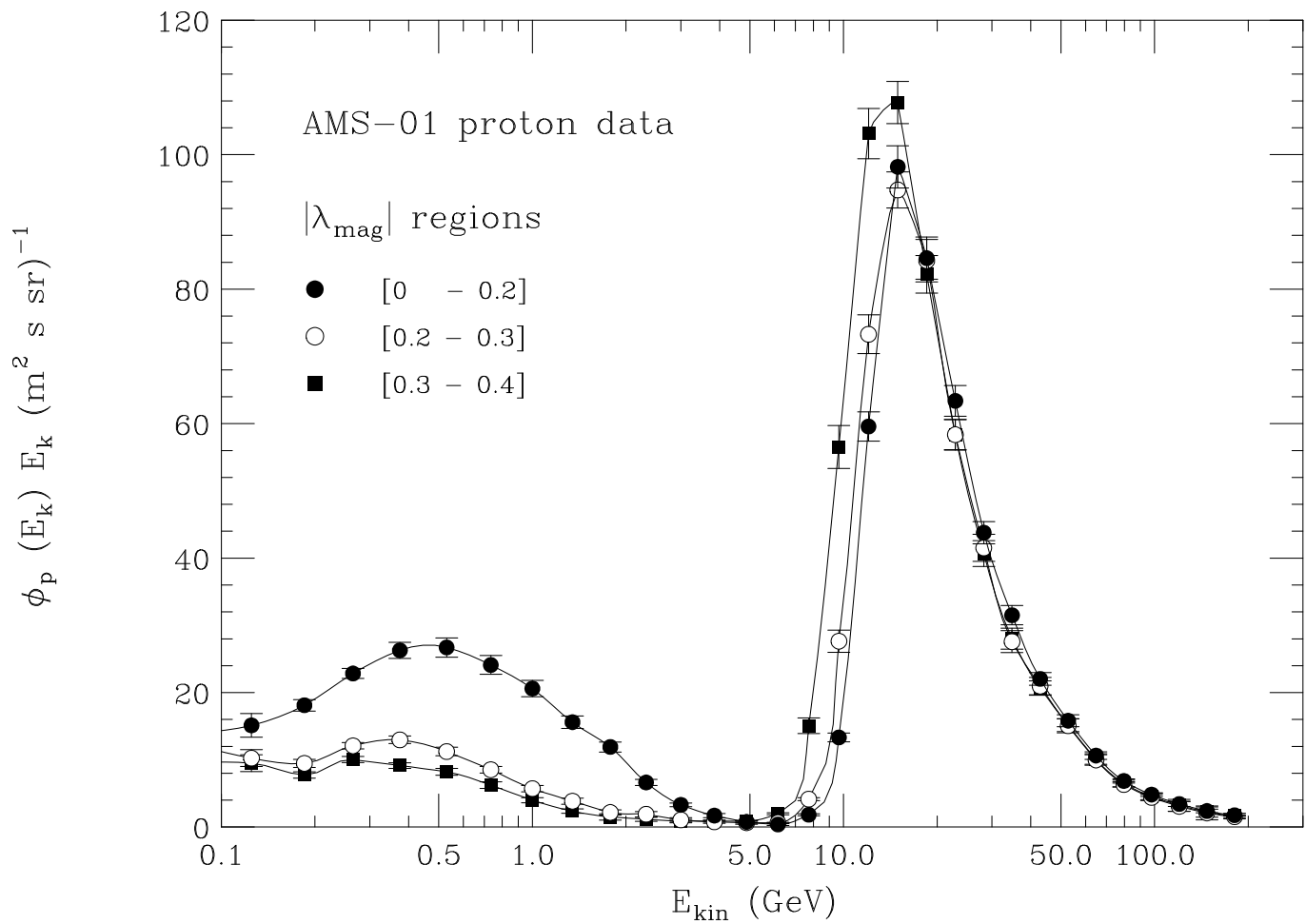,angle=90,height=12.9cm}}
\caption {\small
  Cosmic ray proton flux  measurements  of AMS
 \protect\cite{AMS-protons}   averaged  for  a detector position
in three regions  of  magnetic  latitude.
\label{fig:ams2}  }
\end{figure}

The Anti Matter Spectrometer (AMS)  collaboration  
has recently published  the results on
 new measurements of the fluxes of cosmic  ray protons
\cite{AMS-protons}, 
electrons  and
positrons  \cite{AMS-leptons} and Helium 
\cite {AMS-helium}  performed
during a  ten days flight (STS--91)  of the space  shuttle in june 1998.
These   measurements  have  revealed the existence   of  
significant fluxes  of 
particles   traveling on ``forbidden  trajectories'',
that is   with   momenta below the  calculated  geomagnetic
cutoff.  These fluxes are refered to as  ``second spectra'' in the
AMS papers.
Examples  of the energy distributions   of $p$  measured  by AMS
are shown in fig.~\ref{fig:ams1} and~\ref{fig:ams2}.
The  second spectra  fluxes exhibit  a number of striking
properties:
\begin{enumerate}
\item The ratio ${}^3$He/${}^4$He   is of order $\sim 10$, 
two order of magnitude  larger than  the ratio found in the
primary flux.

\item The ratio $e^+/e^-$  is of order $\sim 4$,   to be compared with
a  ratio  $\sim 0.10$ for  the primary fluxes.

\item 
The past and future trajectory  of each observed second spectrum 
particle can be  calculated  integrating the classical  equations of motion
of a charged particle in a detailed map of the geomagnetic field.
All second--spectra particles appear to have origin  
in the  Earth's  atmosphere, and  to have trajectories that end in the
atmosphere.
The calculated flight time 
between the  estimated origin and  absorption points
has  a broad  distribution extending from  $\sim 10^{-2}$ to $\sim 10$~seconds.
This time is  much shorter  than the  typical confinement 
time of particles in the radiation belts, however
the  higher   end of this  range  is also much longer than the 
Earth radius  ($R_\oplus/c \simeq  0.021$~sec).

\item Two different ``classes''  of  particles
with different  properties 
are seen:  ``short'' and ``long--lived''  particles,
depending on the  calculated  time of flight 
($t$ greater or smaller than $\sim 0.3$~seconds).
Long lived particles  account for most ($\sim 70$\%  for $p$) of
the  second spectra fluxes.

\item  Long lived particles appear to  originate   only from
some well defined regions of the Earth's  surface  that are
different and well separated  for  particles  of different
electric charge.
For example, positively charged  particles ($p$, $e^+$ and Helium)
detected in the magnetic equatorial region
appear to  have their origin 
from  points in the same  range of magnetic  latitude
but with longitude
confined in the   approximate  interval: $\varphi \in [120^\circ,300^\circ]$.
On the contrary   the calculated  points of  origin of $e^-$    have longitude
in the complementary interval  $\varphi \in [-60^\circ,120^\circ]$.
In the ``allowed interval'' of longitude the  distributions
of the   creation points have   a  non  trivial  structure.
For  example most  long lived  positively charged particles
  originate  from  the longitude sub--intervals
 $\varphi \in [120^\circ,180^\circ]$   and 
 $\varphi \in [220^\circ,280^\circ]$.

\item Similarly the absorption points of the long lived  particles
are confined  to well defined regions, that are well separated
for positively  and negatively charged particles.
The ``sink''  regions  for positively charged particles
approximately coincide  with the ``source'' regions  of
negatively charged  particles  and viceversa.

\item For  ``short lived''  particles the  positions  of 
the estimated   origin and
absorption  points do not exhibit the  interesting patterns  described  above.

\item The $e^+/e^-$ ratio  is $\sim 4$  for  long--lived
and  $\sim 2$ for  short--lived particles.

\item The intensity of the  second spectra  fluxes is  large.
For  protons  in the magnetic equatorial region $|\lmag| < 0.2$
($\lmag$ is the magnetic  latitude)
it is of order  $\sim 70$ (40) in units (m$^2$~s~sr)$^{-1}$
for  $E_{\rm kin} > 0.1$ (0.3)~GeV.
This  is  comparable with the intensity of the
primary $p$ flux reaching  the equatorial region
(of order $\sim 100$  in the same units).
\end{enumerate}

The natural candidate mechanism 
as  the source of  the  second spectrum particles  is the  
production of secondary particles in the showers    generated
by cosmic rays  in the atmosphere.  
A  fraction of these  secondary particles, produced
with up--going  directions,  or bent into up--going directions
by the geomagnetic field  can in fact reach high altitude.
These secondary particles 
have  been  known in the literature as the 
components of the ``cosmic ray albedo''
\cite{albedo} (see also  \cite{albedo-HEAT} for  a recent  measurement
in a  balloon   experiment and  additional  references).
In this  work   we will show  that the simple  hypothesis
that   the cosmic  ray albedo is the source of 
the observed proton second--spectra is  in good
qualitative and  quantitative agreement with the AMS data.
This   conclusion  has also   been reached  by  Derome et al. \cite{derome}.
We will also discuss  that 
all the  ``striking'' properties    of the second--spectra
discovered  by AMS  have 
 simple explanations  in terms of the  ``albedo model'',
and while not predicted, can   be simply and naturally  understood
a posteriori.

This work   has  been  also  motivated  by  the importance of  a  detailed
understanding  of the cosmic  ray  (c.r.) fluxes for the 
calculation of the  atmospheric  neutrino  fluxes.
Recent  measurements of these  fluxes by the
Kamiokande \cite{kamioka}, IMB \cite{IMB},  Super--Kamiokande \cite{SK},
MACRO \cite{macro} and Soudan--2 \cite{soudan} detectors
have  given evidence  for the existence of neutrino oscillations
(or possibly some  other form of  new physics  beyond the standard model
 \cite{exotic}).
The  strongest evidence  \cite {SK} comes from  the  observation of  the 
 suppression of the up--going 
$\nu_\mu$  and $\nubar_\mu$ fluxes  with respect
to the down--going ones.
This  can  be interpreted as the ``disappearance''  of  a
fraction of the   $\nu_\mu$  ($\nubar_\mu$)  that  travel long  distances
(comparable to the Earth radius) and     transform   into
$\nu_\tau$ ($\nubar_\tau$) that are nearly ``invisible''  because 
most  of them have  energy below the threshold for 
$\tau$ production.
Down--going  neutrinos are produced  by c.r.   striking the
Earth's atmosphere   near the detector  position, 
 while  up--going neutrinos are produced in  showers 
generated  by c.r.  above  a much larger region of
the Earth in the opposite hemisphere.
It is clear that to   reach the conclusion that  $\nu$  oscillations
are present (or in more detail to estimate  how many neutrinos  ``disappear''
or  change flavor)
one needs to know with sufficient  precision the relative intensity
of the cosmic rays  fluxes   striking the
atmosphere  over  different  positions on the Earth.
Geomagnetic effects  are  important in  determining then intensity
and angular distribution of the  c.r. fluxes   that
reach  the Earth atmosphere, and have to be  described  correctly.
To test the  correctness of  the  treatment of these effects
the  AMS  results  are clearly of great value.
In fact, because of the  inclination of the space shuttle 
orbit  during flight STS-91
(51.7$^\circ$ with respect to the  equator)
 and   the Earth's rotation,
the AMS  detector has measured, essentially
simultaneously,  the cosmic  ray fluxes over  a 
a large  fraction  ($\sim 89$\%) of the  Earth's  surface.

The calculations of the atmospheric  neutrino fluxes,
that have been used in the interpretation of the data
\cite{Bartol,HKKM,fluka}, have assumed 
that the fluxes of   c.r.  particles  below the geomagnetic  cutoff 
are  exactly vanishing.
The  new measurement  tell us that this is  an 
incorrect assumption,   since the  forbidden  trajectories
are populated by second--spectra particles.   It is 
therefore necessary to  investigate  the 
possible impact on  these sub--cutoff  particles  on 
the predicted intensity  and angular distributions
of  the  neutrinos fluxes. 

Our  conclusions are that the AMS  results  give an important
confirmation of the   assumption (of  crucial
importance for the neutrino  flux  calculations) 
that  the cosmic  ray flux in  interplanetary space,
when it is not disturbed by the geomagnetic  effects, is isotropic.
Furthermore the  AMS data tell us  that  the  description of geomagnetic
effects used in recent calculations of the atmospheric
neutrino  fluxes \cite{Bartol,HKKM} are  correct.
The observed   fluxes  of  sub--cutoff particles   represent a
neglibly small correction ($\sim 0.1\%$) to the  neutrino  event  rates.

This  work  is organized as  follows:
in the next  section 
 we will   discuss a montecarlo calculation  of the proton
albedo  fluxes  and compare the  results with the AMS data;
in section 3  we  discuss the results, and illustrate
 how  the observed properties
of  the albedo (or  second--spectra) fluxes can  be  
understood  with simple  qualitative arguments
independently from a  detailed  calculation;
section 4  discusses  the impact of the albedo  fluxes on  the 
estimates of the atmospheric $\nu$ fluxes.  
The last  section  gives a summary and some conclusions.
In an appendix  we  briefly present 
some  general  and well known results about  geomagnetic  effects
that are  used in this work.

\section{Montecarlo  calculation}
\label{sec:montecarlo}

In this  section we will discuss 
a  preliminary montecarlo 
 calculation of  the fluxes of  ``cosmic ray albedo''  protons.
In the  following  an ``albedo'' particle  is a secondary particle  produced
by cosmic  ray interactions in the atmosphere  with a  trajectory  that
reaches  a  minimum altitude  $h \simeq  380$~Km 
(the  approximate  altitude  of the space shuttle during flight STS--91).
This  is a straightforward problem, that is 
essentially identical  to a
(three--dimensional) calculation of the atmospheric  neutrino fluxes.
The montecarlo code  is  in fact the same one  used for
 the  calculation of  the atmospheric  neutrino fluxes described in
  \cite{pl-3d}. It can be described as follows:
\begin{enumerate}
\item An isotropic    flux of primary cosmic  rays
(protons and   heavier nuclei)  is  generated  ``at the top of the
atmosphere''  (a surface with  altitude $h = 100$~Km).
\item To take into account of the geomagnetic  effects,  the  past 
trajectory of each  particle is  studied, and particles on 
forbidden trajectories are rejected
(see the discussion in \ref{subsec:allowed} in the appendix).
The region   with  altitude $h > 100$~Km   is considered as ``vacuum''
except for the presence of a static  magnetic   field
described  by the IGRF  model \cite{IGRF}.

\item Each primary particle   is  then ``forward'' 
traced in  the   ``atmosphere''  (that  is the region with $h < 100$~Km)
taking into account   the presence of the magnetic  field, and 
a variable air density  (described as the average standard US atmosphere).
An  interaction point is  randomly chosen,  
taking into account the  cross section  value and the variable  air density.
A few primary  particles  ($\sim  3$\%)   cross 
again the  injection surface ($h = 100$~Km)  without  interacting and
are discarded. 
\item  The trajectories of all secondary protons
with kinetic  energy  $E_k > 300$~MeV (that is above threshold  for
$\pi$ production) produced in a  primary interaction 
have  been calculated   integrating the equation of  motion 
in a map of the geomagnetic  field. 
 Most secondary protons 
reinteract after  traveling  a  short distance, however
a  few  particles  will  reach high  altitude.
The only difference with  the neutrino flux  calculation  \cite{pl-3d}
is that while in that work  all stable particles that
exited  the ``atmosphere''  (that is the volume
$h \le 100$~Km) were  discarded. 
These particles  are  now the  the object
under  study, and their 
trajectories  are   calculated   in detail, 
following   the motion 
 until they   reaches a  radius
$r = 10~R_\oplus$  (when the particle is  considered as ``free''),
or interact   in the atmosphere.

\item  To estimate the flux  associated with these
``albedo''  particles  we have 
considered a ``detector  surface''  at a constant  altitude
$h = 380$~Km.  Each time a  proton
crosses the detector surface  (up--going or down--going)
a ``hit'' is recorded, and a flux  is  estimated from the number of hits.
\end{enumerate}
 
\begin{figure} [t]
\centerline{\psfig{figure=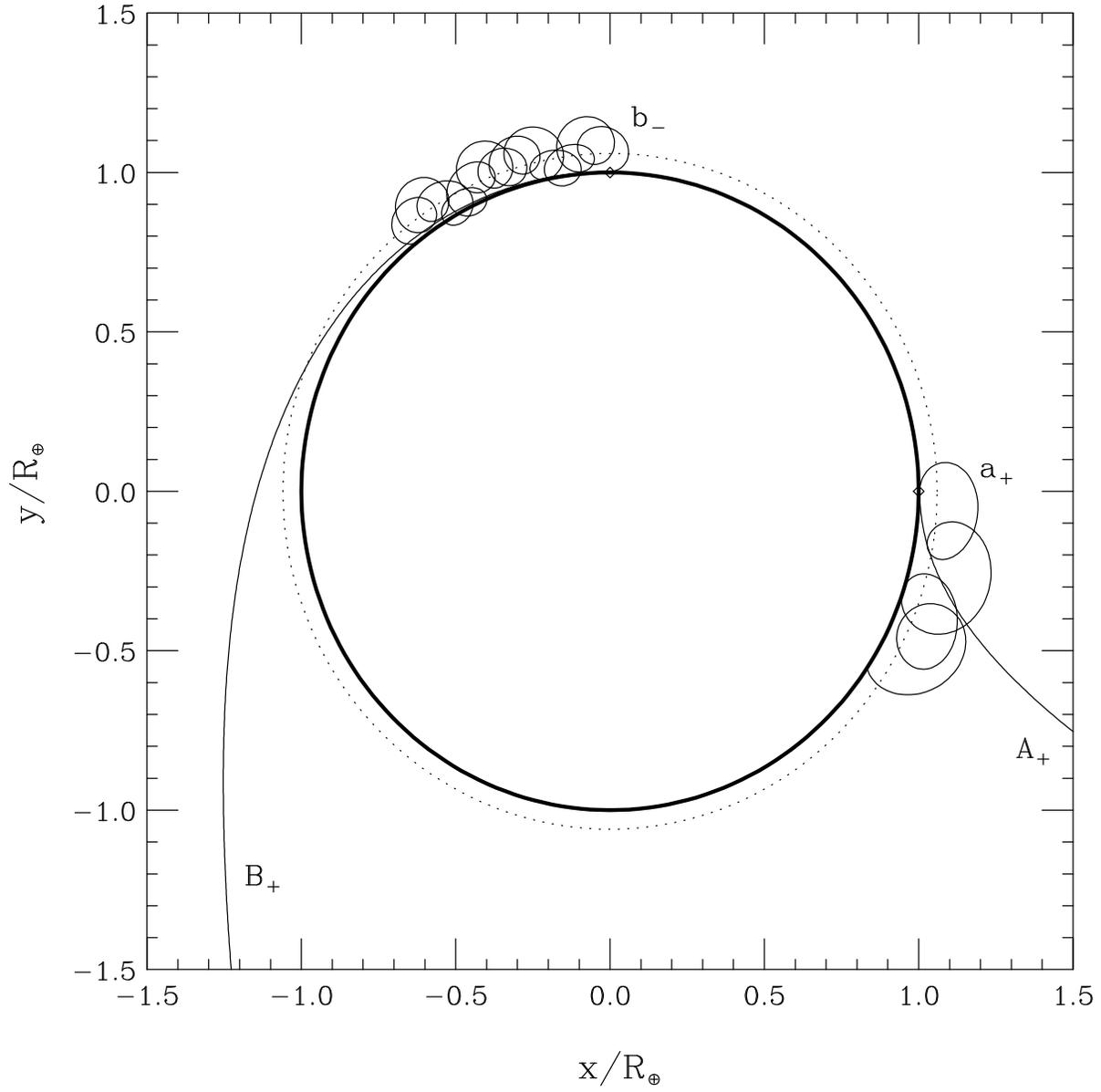,angle=90.,height=16.cm}}
\caption {\small  
The   Earth's equatorial  plane with the projections of the 
trajectories of 4 particles  with rigidities:
$R_A = 30$~GV,
$R_a = 5$~GV,
$R_B = 80$~GV and
$R_b = -4.5$~GV.
The  diamonds  indicate the final (starting) point of the trajectories for 
particles $A$ and $B$ ($a$ and $b$), these  final (starting)  points
are  on the equator, and  have  longitude $0^\circ$
and  $90^\circ$  and   altitude 20~Km. 
 The  dotted  line indicates the altitude
of the space shuttle orbit.
\label{fig:geom}
}
\end{figure}

Some  examples of trajectories of secondary particles
that  reach the altitude of the space shuttle orbit are seen
in fig.\ref{fig:geom},~\ref{fig:traj1},~\ref{fig:traj2} and~\ref{fig:ex1}.
A discussion of the properties of  these trajectories
will be made in the next section.

As an illustration of the calculation,
in fig.~\ref{fig:energy} we  show   (with   correct relative  normalizations)
the assumed energy spectra  of primary protons in interplanetary space,
the energy spectra of the 
protons   that interact  in the atmosphere in  the
equatorial  region $|\sin \lambda_{\rm mag}| < 0.4$,  and the energy spectra
of  all secondary protons  generated in the interactions in the same region.
Comparing the first two   histograms  the effect
 of the geomagnetic field in suppressing the flux of primary  low rigidity
particles can be easily  seen.

In fig.~\ref{fig:zeni} we show  
(with correct  relative normalization) the zenith angle  distribution
of all   nucleons that interact in the  magnetic  equatorial  region
($|\sin \lmag| < 0.4$), the zenith angle  of all  nucleons that are
the source  of  an albedo proton, and  the zenith  angle  of albedo
protons  at the production  point.
As it is  intuitive,   quasi--horizontal  trajectories, 
and ``grazing''  trajectories
that   do not intersect the surface of the  Earth  play  an essential
role as   the source of albedo particles.

In fig.~\ref{fig:azi} we show 
(for  the region $|\sin\lmag| < 0.4$)
the azimuth angle  distribution
of  interacting  nucleons,  
of the nucleons that are the source of albedo $p$, 
and of  albedo protons  at the  production point.
Most of the interacting   primary particles 
are traveling   toward the east.
This is   the  well known east--west  asymmetry  \cite{Rossi,johnson,alvarez},
whose detection   in the 1930's allowed  to determine  that  most c.r.  have
positive electric  charge.
The  enhancement of  the east--west effect for albedo particles is 
qualitatively easy to understand  
because the effect  of the geomagnetic  field
on positively  charged   particles  going east  (west)  is to    bend them
upward (downward)  (see fig.~\ref{fig:geom} for an illustration).

In fig.~\ref{fig:hh} we show the  the altitude  distribution of the
 all nucleon  first interaction points,  and the altitude distribution
of the production points of the albedo particles.
As it is  simple to understand, most albedo particles  are created
at high altitude.

The time  $t$ associated to  a computer run can be 
calculated as:
\begin{equation}
N_{p, {\rm gen}} = 
t~A_{\rm gen}~ \pi~\int_{E_{\rm min}}^\infty dE ~\phi_p(E)
\end{equation}
where $N_{p, {\rm gen}}$ is the total  number of primary proton
simulated (including  those rejected because of  geomagnetic  effects),
 $A_{\rm gen} = 4 \pi~R_{\rm gen}^2$  is the 
area  over which the primary particles  are  generated, and
$\phi_p(E)$ is the  isotropic (interplanetary space) flux
of primary protons.

Within  the  statistical  uncertainties of the Montecarlo the 
upgoing  and down--going   fluxes of  secondary particles
are  identical. To estimate the flux,
 because of the poor montecarlo statistics
we have  approximated  the  flux of albedo particle at the 
altitude of the AMS  orbit as isotropic.
For  an  isotropic  flux  $\Phi_{\rm iso}$  the rate  of particle crossings
a unit area  detector surface is:
\begin{equation}
 n_{\rm cross} = 2\pi~\Phi_{\rm iso} ~c \beta
\end{equation}
where $\beta$  is the particle  velocity 
and a  factor  of two takes  into  account up--going and 
down--going  particles.
We have then  estimated the   flux  of  albedo particles  observable at
the altitude of the AMS  orbit  as:
\begin{equation}
\Phi_{\rm albedo} = {N_{\rm cross} \over 2\pi ~c\beta\, t~A_{\rm det} } =
{1 \over 2 \,c \;\langle \beta \rangle } 
~{N_{\rm cross} \over N_{p, {\rm gen} } }
~{A_{\rm gen} \over A_{\rm det} } ~\Phi_{p, {\rm primary}}
\end{equation}
where $N_{\rm cross}$ is the total number of recorded crossing 
of  a ``detector'' surface of total area $A_{\rm det}$.
For  a ``detector'' over the region $|\lmag| < 0.2$, the resulting flux
is  shown in fig.~\ref{fig:flux}.
In the same  figure we also show  the flux  of vertical  
protons  (selected in a cone  of zenith angle $\theta < 32^\circ$)
averaged over all positions  in the  selected  region.
The result is compared with the  AMS  measurement in the same 
geographical  region.
The  agreement is not perfect  but reasonable, and  the order of  magnitude
of the  flux  and qualitative features of the  measurement are
well reproduced. 

In fig.~\ref{fig:n0}  we show   the distribution of the number
of crossings  that all ``detected'' particles.
Particles with one  crossing are in part ($\sim 40$\%)
protons  produced  with sufficiently high rigidity, so that they
can escape to infinity  never ``returning'' to the Earth',
and in part ($\sim 60$\%) particles that   that have a  second  
(or more) crossings outside the selected equatorial region.
The most probable   situation  for an albedo  particle  is to have  
two  crossings. This  corresponds  to  a proton  that
is  generated   in the atmosphere,  goes to high altitude  and  returns
close to the surface of the Earth where it is absorbed.
We can see that there is also a significant  probability of
4, 6, and    more crossing,  up to more than  $10^3$
(an odd number of crossings correspond to a situation 
with at least one crossing outside
the equatorial region).

The contribution  of an albedo particle to the flux  is  proportional
to  the number of crossings   of the detector surface.
In fig.~\ref{fig:n1}  we show   an  histogram of the
relative  contribution of particles with $N_{\rm cross}$ crossings
to the estimated  flux.  One can see that albedo 
 protons  with  long flight--paths
even  if  they  constitute a small  fraction of the total  number 
particles contribute most  of the observed  sub--cutoff  flux.

In fig.~\ref{fig:points}  we show the positions of creation and absorption
of  ``long--lived'' particles (a  flight time  longer than  0.3~seconds,
according to the definition of  AMS).  Histograms  of the 
longitude  distribution  of the points is shown in fig.~\ref{fig:longitude}.
The  remarkable   features of the distribution are in reasonable 
agreeement with  the experimental results of AMS.

The results for the $p$   albedo fluxes
in  other  regions of magnetic  latitude  have the same  level 
of agreement  with the AMS  data.

\section{Qualitative discussion}
\label{sec:qualitative}
In this  section we  want  to   illustrate how 
all the most interesting   features  of the second spectra
observed by AMS   (listed in the  introduction)
can  be  naturally   explained
in the  framework of a model  where these spectra are composed
of albedo particles.

\subsection {Origin of the particles}
A first remark is  that 
when the  past trajectories of the  particles of the second  spectra 
 are   calculated,  it is 
found that all 
detected  particles originate  in the  deep atmosphere.
This     result is not  trivial.
Second spectrum  particles   are  by  definition 
on forbidden  trajectories, that is 
they do not come  from  ``infinity'', 
however this  does {\em not}  a priori  imply 
that their past  trajectory  will  cross   the Earth's surface. 
In fact  most  trajectories of  particles  in the radiation  belts 
\cite{Walt} do  not  originate   directly  from the Earth's   atmosphere 
and  have  past trajectories  that  remain   confined
at high altitude.
The creation point of these particles   are inside the belt volume 
where the  residual air density is  very low, and 
for most particles the production mechanism is  
neutron decay. 
The second  spectra on the other hand are   produced inside the
atmosphere, as  cosmic ray albedo.

\subsection {$^3$He/$^4$He ratio}
The  very high  $^3$He/$^4$He ratio  observed  in the
second  spectrum is  a  very simple phenomenon
to understand qualitatively.
The key  facts are that: (i)  there is  a  large flux
of  primary $^4$He in the primary radiation; and (ii)  that there is
a large   fragmentation cross section for the process
${}^4{\rm He} + {\cal A} \to {}^3{\rm Helium} + X$ (where 
${\cal A}$ is an  ``air'' nucleus) 
that accounts for $\sim 30\%$ of the inelastic cross section.
Note that the presence of a  fraction  of $\sim 15\%$ of $^3$He
in the primary radiation,
much larger than the universal  isotopic  abundances, 
 is the result of the  fragmentation  of a  fraction of the 
accelerated Helium   during propagation  in the   interstellar medium.
In first approximation, in the fragmentation process the momentum  per nucleon
remain approximately unchanged. Therefore   after   the breaking--up process
a  $^{4}$He nucleus   of  momentum $p_4$ (and  rigidity
$R_4 = p_4/2$)   produces   $^{3}$He  fragments  of
momentum  $p_3 \sim  {3 \over 4} p_4$ 
 and   rigidity  $R_3  =  p_3/Z  \sim {3 \over 4}~R_4$.
The  key result   is  that the  rigidity of the $^3$He fragment
is smaller  that  the rigidity of the projectile, and   most  $^3$He
fragments   produced   by primary  $^4$He  with rigidity
less that ${4 \over 3}$ times the   cutoff for the 
location and direction considered will be below  cutoff
and potentially trapped.
 A fraction of these   fragments (especially those   generated 
horizontal, east--going parents)   will be injected into
albedo particle trajectories.

In the fragmentation  of a 
$^4$He  nucleus, there are  approximately  equal probabilities
to generate  
$^3$He  and  Tritium  fragments (that can be considered  as stable
for  the  relevant short time scale of this problem).
 However this  does  {\em not}
imply that the populations  of the two nuclear 
species in the albedo spectra  are
approximately equal.  Since  a  Tritium fragment has only one unit
of electric  charge, the relation  between 
  its rigidity  and  the rigidity of the 
primary $^4$He  particle is $R_T \simeq {3 \over 2} R_4$.
In this case 
the fragment   rigidity is larger than the  primary particle one,
and  most Tritium fragments do not remain confined.

Helium--4   nuclei can be injected into  trapped  orbits  when a
primary  $^4$He    particle loses  energy in an elastic scattering,
or when a $^4$He   fragment is produced in the interaction
of  heavier nuclei (such as $^{12}$C or $^{16}$O).
These  heavier  nuclear species are less abundant  than Helium in  the 
primary flux, moreover  since   $^4$He  and 
the most common  heavier nuclei
have the same    $A/Z$ ratio, therefore
the  rigidity of an  helium--4  fragment 
will be (most of the times)  equal  to the  rigidity of the
parent primary particle, that is   above
the geomagnetic  cutoff. From these  arguments  one can
reach the conclusion of a strong  suppression for the injection of 
Helium--4 into trapped
trajectories.

\subsection{Positron/electron ratio  in the albedo fluxes.}
The   qualitative 
reason for the  high  $e^+/e^-$ ratio   observed in the subcutoff fluxes
is illustrated in fig.~\ref{fig:geom}
that shows a map of  the Earth's  geographical equatorial plane
with the calculated  trajectories of  four   charged particles.
Particle A is a primary  proton 
with  momentum   $p_A = 30$~GeV, that
interacts  at the   point indicated by a   small diamond, where it 
produced a secondary  particle $a$ 
with  positive unit charge  and  momentum $p_a = 5$~GeV that 
crosses  several times  the altitude $h = 380$~km, before  being  reabsorbed
in the atmosphere.
Both  primary and secondary particle are east--going, and  
reversing the electric charge  of  $a$  would  result in the immediate 
absorption of the particle.
Particle B   has also  unit  positive  electric charge,
and  momentum   $p_B = 80$~GeV, and reaches the  Earth  atmosphere
traveling on a west--going on a  quasi--horizontal trajectory.
At the interaction point 
a secondary   particle $b$  with negative unit charge
 and  momentum $p_b = 4.5$~GeV  is produced, that  again  is injected in
a trajectory that crosses several times the altitude  $h=380$~Km  before
being absorbed.  Note that  if a particle  is  west--going, it can be injected
into the albedo  flux  only if it is negatively charged.

The  figure  illustrates  three fundamental  points:
\begin{enumerate}
\item  albedo particles 
are most  easily produced
with initial zenith angle $\sim 90^\circ$ (approximately horizontally),
and azimuth angle  pointing  east (for 
 positively charged particles) and  toward west
(for negatively charged ones).
\item The directions   of the  primary and  secondary particles
are  correlated  because of momentum  conservation.
\item  The  rate of east--going primary particles 
is  larger than the rate of west going ones.
The  origin of  this asymmetry (the celebrated  east--west effect
\cite{Rossi,johnson,alvarez}) can  be   easily  understood  looking
at fig.~\ref{fig:geom}, where one can
see  that  positively charged primary  particles
can reach the  Earth equator  from  an horizontal  west--going  direction
only   if they have   a  gyroradius larger that  $R_\oplus$
(this correspond to a rigidity $p/Z \aprge 60$~GV),  while 
the rigidity cutoff for  east--going particles is  much lower
($p/Z \aprge 11$~GV). 
\end{enumerate}
Since the production of electrons and positrons in
hadronic  showers is approximately equal, 
it is now  easy to reach the conclusion that
the injection of albedo  positrons,
(mostly produced in the showers of 
nearly horizontal   east--going primary particles)
is significantly larger that the  injection of electrons
(mostly produced in the showers of 
less numerous west--going primary particles).

A remarkable property of the  $e^+/e^-$ ratio  
measured  by AMS \cite{AMS-leptons},  is   
the fact that the ratio
for  short lived  ($e^+/e^- \sim 2$) and  long lived
($e^+/e^- \sim  4$) particles   differ
by  a factor of approximately two, with only  a  small  energy dependence
for  $E \aprle 1$~GeV.
A qualitative explanation for this  interesting phenomenon   will be given in 
section~\ref{subsec:e+/e-long}.

\subsection{The   longitude   distribution  of the points  of origin}
The long lived  positively charged particle
observed in the magnetic  equatorial  region have their
points of origin  in the longitude range 
$\phi \in [120^\circ,300^\circ]$  while  negatively charged
particles  have their  origin in the  complementary 
longitude interval $\phi \in [-60^\circ,120^\circ]$.
This can  be  understood  immediately on  the basis of three
simple  observations.
\begin{enumerate}
\item Approximating the geomagnetic  field
as  a dipole,  one  finds that the dipole
is not only ``tilted'', that is it  with an  axis   not
parallel to the Earth's  rotation axis, but is also
``offset''  that is  the  dipole   center does not coincide with the Earth's 
geometrical center.
\item  The motion of charged particles    confined  to the equatorial plane
of a magnetic dipole can be analysed as a 
``gyration''  around  a guiding center  that   drifts   uniformly
in longitude remaining at a constant distance from the dipole center
(see  fig.~\ref{fig:ex1} and the discussion in \ref{subsec:trapped}).
\item  Positively  charged particles 
drift westward  (toward  decreasing longitude)  
while  negatively charged particles  drift eastward 
(toward   increasing  longitude).
\end{enumerate}
A scheme  of the  drift of  particle   in the equatorial  plane
of an  offset dipole model  is  shown in fig.~\ref{fig:eccentric}.
It is  simple  to  see  that
positively  particles  can be injected into 
albedo trajectories  only if produced  in one hemisphere, and
are absorbed in the opposite hemisphere, while   the  opposite
happens for  negatively charged  particles, since the
longitude  drift of the guiding  center  of the trajectory
traveling at a constant distance from the dipole  center
has  a  variable  altitude, that begins  to decrease for
particles created in the ``forbidden'' hemisphere, or to increase
for  particles created in the  ``allowed'' one.
The argument  can be easily extended to 
the general case of  of trapped trajectories, when 
the longitude 
drift is accompanied  by  an oscillation or  ``bouncing motion'' 
 along  the  field  lines  between  symmetric mirror  points.
The  altitude of  the 
guiding  center of the trajectory  has  minima 
(where the particle  has the highest chance of being absorbed) 
at the  mirror points   that have  a constant  distance from
the dipole center.
In an offset dipole field the  altitude of  the mirror points change
with the longitude drift.
This mechanism can also  be described 
in a more  general,  elegant 
 and rigorous way  making use of  the concept
of magnetic  shells (see section~\ref{subsec:shells} in the appendix),
and   considering the  intersections of the shells  with the Earth's surface.

For  a description  of the  consequences of the dipole offset,
it is convenient  to  ``shift''  the origin of the longitude 
defining:
\begin{equation}
\longa= \varphi - \varphi^*
\end{equation}
where  $\varphi^* \simeq 120^\circ$ is 
the  longitude of  the dipole  center seen from the Earth's center
(or also the longitude of the point of strongest field  for a fixed
latitude), 
and  use the  convention that the ``shifted  longitude''  is defined
in the interval $[-\pi,\pi]$.
Then it is simple to see
that the source and
sink regions for  positively and  negatively particles 
are  confined in longitude:
\begin{eqnarray}
& & [{\rm Source}]_+  \simeq [{\rm Sink}]_- \simeq
\{\longa > 0 \} \\
& & [{\rm Source}]_-  \simeq [{\rm Sink}]_+ \simeq
\{ \longa < 0 \}
\end{eqnarray}

In the eccentric dipole model
it is  simple to  predict 
an approximate  one--to--one correspondence
between the creation  and absorption points of long--lived particles.
An albedo  positive particle  created at longitude
  $\longa^i_+$ 
(with $\longa^i_+ > 0$)   will be absorbed
either  ``soon''  or   will  drift  for a ``long''  time
clock--wise, that is 
toward decreasing   $\varphi$. 
Let us  consider for  simplicity the motion  of particles  confined
in the  magnetic  equatorial plane. 
 During the  drift  the    guiding center  of the particle trajectory
remains  at a constant distance  from the dipole center,
therefore because of the dipole offset,  the  distance $r$
from the Earth center  changes.
For particles  produced in the ``allowed''  longitude range
$r$ starts  increasing, and  soon
 absorption in the atmosphere becomes  impossible.
The  growth of the altitude of  the  trajectory guiding center 
will continue  until  the shifted longitude is  $\longa \simeq 0$,
 then symmetrically it  start  to decrease.
When the longitude  becomes 
\begin{equation}
 \longa^f_+ \simeq -\longa^i_+
\end{equation}
the particle  will again   be  grazing
the  atmosphere,   and  will be reabsorbed.
Symmetrically a negatively charged albedo  particle created at 
a  point   $\longa^i_- < 0$, 
will either be  quickly absorbed or  drift counter--clockwise
until  it reaches  longitude
\begin{equation}
 \longa^f_-  \simeq -\longa^i_-    .
\end{equation}
In both  cases  the   total  longitude drift is
\begin{equation}
\left | (\Delta  \varphi)_{\rm drift}^\pm \right |
 \simeq  2 \, |\longa_i^\pm| 
= 2 \, |\longa_f^\pm|
\label{eq:drift-interval}
\end{equation}
Note that particles  produced  with 
shifted longitude close to $|\longa| \simeq \pi$ drift
for a  longitude interval  close to $2\pi$  that is 
  nearly an entire  Earth orbit, while particles
with  longitude close  to $\longa \simeq 0$  will drift for
a short distance.

This  argument  lead  to  the prediction of a
simple relation  between the 
time  of  flight of albedo particles, their position of  creation
and the  momentum.
The  angular    velocity   of  the  drift motion
was  estimated in equation (\ref{eq:omega-drift}).
Using that  result we can  deduce 
that for  both positively and
  negatively charged particles:
\begin{equation}
t \simeq {|(\Delta \varphi)_{\rm drift}| \over \Omega_{\rm drift} }
\simeq 1.8~ { |\longa_i| \over \beta \, p({\rm GeV})}~{\rm sec} 
\simeq 1.8~ { |\longa_f| \over \beta \, p({\rm GeV})}~{\rm sec}.
\label{eq:diagonal}
\end{equation}
This is  a   remarkable relation   between  three quantities:
the momentum of a second spectrum particle, its  calculated
time of  flight, and   the   estimated  longitude   of the 
creation  (or absorption) point. 
Any choice of a pair  among these three quantities, allows to predict
the third one.
This  relation is verified by the  AMS data (see  figure~6  in 
 \cite{AMS-leptons}).

A  simple  but important prediction 
 of the ``eccentric dipole model''  is that
 it is  difficult\footnote{Nonetheless this is not
impossible, especially for  particles  created 
at large   magnetic  latitude. An example is show in fig.~\ref{fig:traj2}.}
 for a particle to perform  an entire
``drift''  revolution   around the Earth, as can be seen with simple
geometrical  considerations.
This   results in a simple prediction for the
longest  time of flight of  albedo particles: 
\begin{equation}
t_{\rm max}  \simeq {2\pi \over \Omega_{\rm drift} } \simeq
5~ {Z \over \beta \; p ({\rm GeV}) }~{\rm sec}.
\end{equation}

\subsection{Longitude  dependence of the intensity of the albedo spectra}
The AMS  collaboration has  presented   its results on the 
cosmic  ray spectra   for  different  intervals in 
magnetic latitude  for the  detector  position, integrating
over  the  detector  longitude.
However,
since  the  observation of the second spectra    has given evidence
of  striking patterns in the  longitude  distribution of the
creation points  of  the second spectra particles,
it is  natural  to   investigate  the possible 
dependence of the flux intensity on the detector  longitude.
In the montecarlo study described in  section~\ref{sec:montecarlo},
this  dependence has  been  calculated
(see  fig.~\ref{fig:long1})  obtaining a 
non trivial   dependence.

For a qualitative understanding let us consider a
a detector at a position  with shifted longitude   $\longa_{\rm det}$. 
From the results obtained in the 
previous  subsection we  can
infer  that   when   $\longa_{\rm det} >0$
the only observable long--lived particles are  those produced in the
longitude  interval:
\begin{equation} 
\longa^i_+  \in [ \longa_{\rm det}, \pi]
\end {equation}
if   positively charged, and 
\begin{equation} 
\longa^i_-  \in [ -\pi ,  -\longa_{\rm det}]
\end {equation}
if  negatively charged.
Note that the size of   the  two  visible  regions  is :
\begin {equation}
|(\Delta \longa)^+_{\rm visible}| = 
|(\Delta \longa)^-_{\rm visible}| =  \pi - |\longa_{\rm det}|
\end{equation}
is equal  for  both positively and  negatively charged  particles, 
and strongly depends  on the detector  position,    being
maximum  for a detector at $\longa_{\rm det} \simeq 0$,  when  the 
entire production regions  of  both  positive  and   negative particles
is  visible, and  vanishingly  small for  a detector
at $|\longa_{\rm det}| \simeq \pi$.

This  argument seems to   imply that the  intensity
 of the fluxes  of  long--lived albedo particles
are  linear in $|\longa_{\rm det}|$, however 
in this discussion we have not yet  taken  into account
the altitude of  the detector, that  also plays an important  role.
The  altitude of the   guiding center of the trapped albedo particles
also  depend  on the longitude,  it is lowest at the creation 
and  absorption points 
  and it is  highest  at   $\longa \simeq 0$.
Therefore for a detector at a fixed  altitude
($h \simeq 400$~Km  in the case of AMS), only a fraction  of 
the albedo flux will be visible,  some  part  of it being
too low and some  part being too high.

The combination of these arguments: the ``visible   longitude horizon'',
and the ``altitude  of the guiding  center'', 
 can explain the structure of the  numerical  results
 shown in fig.~\ref{fig:long1}
that shows 
a minimum at the longitude  $|\longa_{\rm det}| \simeq \pi$
(corresponding to $\varphi_{\rm det}  \simeq 300^\circ$) and
two   maxima  at longitudes  70$^\circ$ and $150^\circ$, 
placed approximately symmetrically  to the sides of  the  point 
$\longa_{\rm det} \simeq 0$. 
The gyroradius  of  the 
particles   (see equation (\ref{eq:gyro}) is  not  negligiby small,
and also  plays an important role  in determining  which particles 
 are ``visible''.
Therefore the longitude dependence  of the flux,  has  different  shapes
for different momenta.

The  range of  longitude
where we predict the {\em  lowest} intensity  of the albedo fluxes
corresponds to 
the region closest to the south atlantic  anomaly,
where it is well known  that the flux  of trapped particles
is  extremely intense.  This appears  as  a 
paradox, but it can  be  naturally explained.
In  fact the existence of the south atlantic  anomaly 
and the patterns observed by AMS  for  the second spectra fluxes
originate from the same  cause, namely the 
offset of the dipolar component of the geomagnetic  field.
One  consequence  of the   offset
is that the ``magnetic bottle''
that  contains  charged particles in the inner  Van Allen belt
is  not  symmetric with respect to the Earth center.
The south atlantic anomaly corresponds to the
region where a  tip of the ``bottle''  is  closest
to  the Earth surface, descending to an altitude of 
few  hundred kilometers    over an area  of South America
and the south atlantic ocean,  while  on the other side of the Earth
it remains  above $\sim 1200$~Km.
The   existence of the  ``allowed'' and ``forbidden'' hemisphere
for the  production   of long lived  albedo particles
can be understood observing   that
the equatorial  region of  some 
magnetic  shells (see section \ref{subsec:shells})  will 
intersect the  Earth  surface.
Positive  particles  are  created  at one  intersection  between
a  magnetic shell and  the  Earth  surface,  
and drift to be absorbed  to the other  interesection (and viceversa.
for negatively charged particles). 
The   particles are  observable when  the  magnetic  shell 
over  which their guiding center is  moving is close to 
the  space shuttle orbit  altitude\footnote{
Because of the finite gyroradius of the trapped  particles
($a \propto p_\perp$)
 the region where they 
are  observable   depends  on the  momentum.}.
The longitude of the subatlantic  anomaly, is roughly the longitude
at which the   equatorial  region of the  magnetic  shells is
closer to the Earth's  surface.
It follows that 
when the detector is  at the  longitude  of the
anomaly, the  observable 
flux  of sub--cutoff particles   is suppressed, since it is  sitting
on a shell that has   no intersection with the atmosphere 
along its equatorial  region.

\subsection{Long lived and short lived  particles}
In the AMS  papers  the  group  makes a distinction about two classes
of particles, ``long--lived''  and ``short--lived''.
The arguments presented  above 
can  be used to understand the existence and the   
properties of  these two ``classes'' of particles.

The phenomenological  evidence for the two clases of particles
is  perhaps  most evident in
fig.~6  of  the paper \cite{AMS-leptons} on subcutoff
electrons and positrons.
The figure   shows  a scatter plot of   the time of flight  of
 second--spectrum particles
  detected in  the region $|\lmag| \le 0.3$
versus their  kinetic  energy.
  Some  interesting structures
are  immediately apparent:
\begin{enumerate}
\item   There are two    `horizontal bands'
that  is particles  with  a  time of  flight   
$t \sim  0.03$~sec and $t \sim 0.06$~sec,
and a wide range on   energy.

\item  There is a large  ``gap''  in time of flight.
For example there are   few electrons  or positrons  with energy 
$E_k \simeq 0.3$~GeV  and   time  of flight
in the interval $t \in [0.09,2]$~seconds.

\item  There are some  broad   ``diagonal'' bands.
The two  most   evident  bands    (labeled  as $A$ and $B$,
in the AMS  work \cite{AMS-leptons})
are   centered  around the relations:
$t \sim 1.5/p_{\rm GeV}$~sec  and 
$t \sim 4/p_{\rm GeV}$~sec.
The  particles in   each one of the bands originate in a well defined 
and  distinct region  of the Earth's  surface. 

\item Finally no electrons  or  positrons  have been detected
with a time of flight  longer  than 
$t \sim  5/p({\rm GeV})$~sec
\end{enumerate}

Particles in the  ``horizontal  bands'' are the ``short lived'' ones,
particles in the ``diagonal bands''  are the
``long--lived''  ones.
All these  patterns  have a simple  qualitative  explanation.
The existence of the ``horizontal  bands'' 
is  due to the   ``bouncing''  motion of  trapped  charged particles
around the magnetic equatorial plane.
As  discussed in sec.\ref{subsec:motion},  the motion of  a charged  
particle  in a quasi--dipolar field  can be   decomposed
into   three  components:  a  very  fast  gyration  around
a guiding  center,  a fast 
oscillation (or ``bouncing'')  around the magnetic equatorial plane,
and a slow  drift in longitude.
The ``bouncing period''  of  relativistic  particles
is  momentum independent:  $T_{\rm bounce} \sim 0.06/\beta$~sec
(see  equation (\ref{eq:omega-drift}) and the discussion  in
section~\ref{subsec:trapped}).
Because of the   structure of the field
(see for example  equation (\ref{eq:field-line})),
 the  altitude  of the guiding  center 
of a trapped  trajectory  is  maximum on the equator 
and   minimum at the ``mirror  points''.
Particles  are  clearly created  and absorbed near a mirror  point.

The lowest  ``horizontal bands''  in the  
flight--time  versus  momentum   can  be  understood 
as   due to  particles
that    are never  reflected, that is are 
produced in the northern (or southern)  magnetic  hemisphere  and 
reabsorbed  in the  opposite one,  after a single  crossing of the 
magnetic equatorial  plane.
The  next   ``horizontal band''  is due to
particles that  perform  one  reflection,  that is   they
are produced  and absorbed  in the same  hemisphere    after
one reflection and  two   equatorial  crossings.

While  a particle ``bounces'' up and dow, in magnetic  latitude, 
it is also  drifting in longitude.  
The drift   carries the  altitude  of the mirror points
either lower (for  positively charged particles
with $\longa < 0$  and  negatively charged particles with
$\longa >0$)   leading to the  particle  absorption,  
or (in the  complementary cases) higher.
In the  second   situation, if the particle  has  not  being  absorbed 
after the first  two reflections,
 it  has a good chance  to drift  for a long time,
until    the altitude of the mirror points 
returns  to roughly the  initial  level,
at  longitude $\longa_f = - \longa_i$.
The  existence of the ``diagonal  bands'' is  simply the  consequence of 
equation (\ref{eq:diagonal});  the  bands are ``diagonal''
because the     angular velocity of the longitude drift
is  $\Omega_{\rm drift} \propto p$  as  discussed
in section~\ref{subsec:motion}.

In a ``nutshell'':  short lived  particle  are albedo particles
that have only zero or  one   reflection at a mirror point.
If  a particle  manages to have at least  two reflections,
and is  produced  in  the ``allowed''   hemisphere
for  its  electric  charge, 
it  has  then a  good chance to have  a long trajectory
with many bounces.

The argument that we have outlined  here    does  not unexplain
why there are  well defined  ``bands''  of  particles,
or in other words  why    the injection of particles
in  the long  lived  trajectories   is  more likely from 
some regions  than from  other ones.
A  qualitative explanation will be given in the next subsection.

\subsection{Structure in the longitude distribution of the creation points}

In the  previous  subsections, we  have shown that 
a description of the geomagnetic  field  as 
a  tilted  and offset dipole   is  sufficient to understand
qualitatively   several important 
properties  of the  second  spectra fluxes
such as: (i) the existence of two  rather 
well separated classes of ``long''  and    ``short''  lived particles;
(ii)  the  fact that long lived particles 
with positive or negative electric charge
are produced in opposite  hemispheres;
(iii)  the  existence of a simple  relation between the momentum,
 the  longitude of the  creation  (or absorption)  point 
and the  time of  flight of a long--lived ``second spectrum'' particle.

However the  ``eccentric  dipole''  description of the geomagnetic field
predicts a smooth    distributions for  the longitude
of  the  production and  absorption  points  of  the albedo particles.
This  is  not  supported by the data, 
that  show  that   it is  easy to  produced  albedo particles
from  some  regions of the allowed
hemisphere, and  more difficult from others.  
In fact the longitude distribution  of the second spectra particles
clearly exhibits  two   maxima.
These effects   are  reproduced with a Montecarlo  calculation
using a detailed  map  of the geomagnetic  field
(see fig.~\ref{fig:longitude})
that includes  higher  order terms in a multipole  expansion.
It is  however instructive  to understand qualitatively how  the
observed structures  arise.

The argument    that is perhaps  most  suitable
for  a qualitative  understanding  of the    structure  in the 
distribution in longitude  of the production points, is
based on a discussion of  the  ``bouncing''  motion of the
trapped  particles (see section~\ref{subsec:motion}).
Trapped  charged particle
trajectories  in a dipole  field  can be analysed 
as  a   very fast ``gyration''  around  a  guiding  center  that moves
oscillating  along  the field lines between symmetric  mirror points,
and drifting slowly in  longitude.
The same  qualitative structure  of the motion exists  also for
a non exactly dipolar  field. 
The ``bouncing'' motion  is  possible 
only   for  sufficiently small amplitudes, 
so that  both  mirror  points are  at sufficiently high
altitude. A  necessary condition is  obviously that they have
radius $r > R_\oplus$.
Note that since the  drift  frequency   $\Omega_{\rm drift}$ is much
smaller  than the bouncing one $\Omega_{\rm  bounce}$ 
(see equations (\ref{eq:omega-drift})  and  (\ref{eq:omega-bounce})),
it  is a good approximation to  study the bouncing  motion
neglecting the drift. 
In this way the  guiding center of a particle trajectory 
can be  seen  as an oscillation along a  particular field  line. 
The center of the oscillation is the 
 the point  along  the field  line where the magnetic
field is  minimum   (the ``equator'' point $E$  on  the line), while  the 
``mirror'' points $M_1$ and $M_2$   have equal 
values  of  $|\vec{B}|$.
 The amplitude of the oscillation is in 
a one--to--one  correspondence with the pitch angle of the
particle  at the point $E$.  
For  a particle that at the  minimum field
(or equator) point is  orthogonal to the field
(that is has pitch angle  $\cos \alpha_0 \simeq 0$)
the amplitude  of the 
oscillation  vanishes,  with  increasing   $|\cos \alpha_0|$ the
amplitude  of the oscillation grows.
The condition that  the  mirror  points  are  above 
ground  can be written  (see the discussion in  section \ref{subsec:bouncing})
as:
\begin{equation}
|\cos \alpha_0| \le {\rm min} \left [
\sqrt{ 1 - {B_{\rm min} \over B(G_1)}},
\sqrt{ 1 - {B_{\rm min} \over B(G_2)}}
\right ]
\label{eq:mira}
\end{equation}
where  $B_{\rm min}$ is  the  minimum field  along
the line  and $G_1$ and $G_2$ are the  points  of  intersection  of the  field
line with the ground.

For a centered  dipole  field  the 
minimum field point coincides  with the point 
of   highest $r$  along a line  and is symmetrically placed
between  the ``ground''  points $G_1$ and $G_2$, but this is  not true
in the general  case. As an illustration 
  in fig.~\ref{fig:mirror} we show  (in two  separate panels)
two  geomagnetic  field  lines
calculated using the IGRF 2000 field.
The two lines
have  been   selected so that the  maximum altitude   along
the line is  150~Km (the maximum altitude point  is  labeled  $A$),
with  longitudes  $\varphi \simeq -140^\circ$  and  $\varphi \simeq
170^\circ$ (this  defines  uniquely each field line).
On each field line  we have also indicated  the point  $E$ where the field
has the  minimum value.
It is  simple to see that if  the point $E$ 
(the center of the oscillations) is  displaced  with respect
to the geometrical    center  of the line,  the 
maximum amplitude of  the oscillations is  reduced, and so is the 
range of  possible  pitch angles  of  trapped  particles.
Note also that if the ``equator'' point $E$  ha $r < R_\oplus$
 a  complete oscillation is impossible.

Let us now  consider all field  lines  that have the  point   of maximum
altitude at a  fixed altitude  $h$.  
A particular line in   this  set can be identified  uniquely by the longitude
of this point. For each   line  we can  calculate
using  equation  (\ref{eq:mira})  the allowed range 
of   pitch angles. The allowed  interval of 
 $\cos \alpha_0$   is   proportional to
the solid  angle   available for  the injection of
long lived albedo  particles from  the region  in the atmosphere close
to  the   ``ground''  points of the field  line,
that is  the cone  of initial directions,  for which  an  albedo particle
will be able to  complete  an entire  latitude oscillations, and 
enter a ``long''  trajectory.
The result  of the    calculation
for  the  value  $h = 50$~Km is 
shown in   the top panel of  fig.~\ref{fig:mirror1}.
The   height of  50~Km was chosen as  a representative  value
for the field  lines that are most important for  the injection
of  particles  in the albedo spectra from the magnetic  equatorial  region.
This   value  is  of course somewhat  arbitrary, but 
the  qualitative  features of the figure are
independent  the precise value   $h$.
 It can  be seen that the allowed range in pitch angle 
for the different  field  lines exhibits 
some  clear features:
\begin {enumerate}
\item There is  a range  of longitudes  where  oscillations
are  completely forbidden. In this  region the  equator  point
is  below  sea level.
\item  There  are two   additional minima 
one in the region  $\longa >0$ and one  in the region 
$\longa <0$.
\end{enumerate}
In the botton panel  of fig.~\ref{fig:mirror1} we have  multiplied
 the allowed range of pitch angle   by  the   longitude  drift
of long lived  particles created at  that longitude
(equation (\ref{eq:drift-interval})).
Particles  with longer  drifts  give a larger  
contribution to the observed  sub--cutoff flux.
The features  shown in fig.~\ref{fig:mirror1} 
reproduce   qualitatively the structures  observed in the data.

\subsection {$e^+/e^-$ ratio for  short and long lived particles}
\label{subsec:e+/e-long}
The discussion of the previous  section   can be used to obtain 
a qualitative understanding of the observed difference in the
$e^+/e^-$ ratio  for short--lived  and long--lived particles.
For  short-lived  particles (most of which
are  reflected   only  zero or   one time   in their ``bouncing'' motion)
the  enhancement of the  positive partices can be understood as a consequence
of (i)  the  larger flux of east--going primary particles over 
west--going ones, (ii)  the  correlation in   direction between 
secondary and primary particles,  and (iii) the fact that 
only   positively
(negatively) charged particles   produced with east (west) going directions
can   be injected into  the albedo fluxes.
The  same argument  is of course valid also for  long lived  particles,
however in this  case we must also consider  two conditions
that  are necessary for a particle to be ``long lived'':
\begin{enumerate}
\item the particle must be  created in the  ``allowed hemisphere'',
\item  the particle  must be able to  complete a  ``bouncing'' 
oscillations, that is  it must  be generated in an allowed
cone of directions,  that  corresponds to   an  allowed range
of pitch angles. 
\end{enumerate}
As discussed  before, 
if a  particle  produced in  the ``allowed'' hemisphere 
  manages   to  perform  one complete  oscillation 
(two reflections),  it is  then  likely to 
make  many more,  since the longitude drift ``raises'' the  altitude 
of the mirror  points.

In the  previous  subsection we have   estimated  
``phase space'' available
(that  is  the range of  possible  pitch angles)
 for  the injection  of long lived  
particles  from the equatorial region as  a function of the longitude.
The  results are shown  in fig.~\ref{fig:mirror1}.
From the  figure  it can  be seen 
that the allowed  solid  angle  for
the longitude   interval 
$\longa < 0$  ($\varphi \simeq [-60^\circ, 120^\circ]$)
that is the source of  long lived  negatively charged  particles
is    significantly smaller than
the  allowed  range  in the complementary   interval
(the source of positively charged particles).
This  constitutes an additional  suppression factor for
negatively  charged long lived  particles  that  has to be
combined with the effects   due to   east--west asymmetry if 
the  primary flux. 
The  results is a larger 
$e^+/e^-$ ratio for long lived  particles, in agreement with the observations.

\section{Implications  for atmospheric  neutrinos}

To estimate the possible importance of 
the c.r. albedo  fluxes  for atmospheric  neutrinos, it is useful
 to  convolute the 
measured sub--cutoff fluxes at  the  space shuttle orbit with 
an ``event yield''  for  $\nu$   charged  current  interactions.
For  example, for $\mu$ events  the  yield  is defined as:
\begin{equation}
Y_{\mu} (E_p) =
N_A ~\int  dE_\nu ~ \left [
 { d n_{p\to \nu_\mu} \over dE_\nu} (E_\nu,  E_p) ~\sigma_{\nu_\mu} (E_\nu) +
 { d n_{p \to \nubar_\mu} \over dE_\nu} (E_\nu,  E_p)
 ~\sigma_{\nubar_\mu} (E_\nu)
\right ]
\end{equation}
where $dn_{p \to \nu_\mu}/dE_\nu$   and $dn_{p \to \nubar_\mu}/dE_\nu$
are   the average   number  of  $\nu_\mu$  and $\nubar_\mu$
with energy $E_\nu$  produced
in the shower of a  proton  of energy $E_p$,
$\sigma_{\nu_\mu}$  and $\sigma_{\nubar_\mu}$
are  the cross sections 
for   $\nu_\mu$ and $\nubar_\mu$ charged current interactions \cite{sigmanu}
and $N_A$ is   Avogadro's  number.
In principle the $\nu$ event  yield  depends  not only on the energy
of the primary particle, but also on its zenith angle, since
 decay of mesons  and  muons
(that are the $\nu$ sources) are 
 more  probable in inclined showers (see  for example \cite{pl-3d});
however for low  $E_p$  this dependence is neglibly small,
since  all   secondary products   have  low  momentum, and
unstable particles, because of their  short decay lengths 
 ($\ell_{\rm dec} = c\tau \,p/m$)   decay  rapidly 
with unit   probability.
A calculation of  the  yield  for muon  events
estimated using the  hadronic interaction  model
of the  Bartol model \cite{Bartol}  is  shown  in   fig.~\ref{fig:nu_yield}.
The  yield vanishes 
for 
$E_{k} \aprle  290$~MeV that is the threshold  for 
$\pi^\pm$  (and therefore $\nu$) production
and then grows  rapidly  
with  increasing  energy.

The convolution  of the Bartol   $\nu$ event yield
with  the AMS  proton--flux  measured at
  high and  low magnetic latitudes is shown in fig.~\ref{fig:response}.
The  integral  of this  convolution  is
   $\sim 8.4$~(kton~yr~sr)$^{-1}$, 
 for the    high  magnetic  latitude  region and 
$\sim 5.7$~(kton~yr~sr)$^{-1}$
for the magnetic  equatorial  region.
The  difference  between the two rates represents the maximum
possible  size  of geomagnetic  effects.
In the equatorial region the   contribution due to 
second spectrum  protons, obtained  integrating
for $E_k < 6$~GeV,
is $\simeq 0.009$~(kton~yr~sr)$^{-1}$, that
is 0.16\% of the  total.
The smallness  of the contribution of the 
sub--cutoff particles  to the $\nu$ rate, is  a consequence
of their softness.
In the region considered, at  the altitude  of the 
space shuttle orbit, 
the  sub--cutoff protons
represent $\sim 42\%$ of the particle 
flux,
 but only  $\sim 1.8\%$ of the energy flux, moreover 
only 60\% of the second spectrum  is above the kinematical threshold
for $\nu$ production, and the $\nu$ that are produced
are  soft with low  cross section.

The integration over  solid  angle  of the 
effect  is  non  trivial, 
however we  can  observe (see fig.~\ref{fig:ams1} and~\ref{fig:ams2})
that the albedo spectrum  has   maximum  intensity
in the magnetic  equatorial region,  therefore 
simply multiplying  by $4\pi$  we are    making a
conservative  overestimate.   
The result is  a contribution of 
second spectrum protons  to the $\mu$--like event rate
of 0.11~($\mu$~events)/(kton~year),  that   is of order 
0.1\% of  a measured  rate  of  $\sim 100$~events/(kton~year).

This   estimate,   while  already negligibly small, 
represents an  overestimate of  the effect. In  fact the
albedo flux  that is  observed  at high altitude is enhanced
because  of the  magnetic trapping.
To estimate a  correction factor
to  pass  from  the flux  observed at
high altitude
to  the   flux  that is absorbed in  the atmosphere 
and  is relevant for the production of
 secondary particles (such as neutrinos),
we can  use the results of 
the montecarlo  calculation described in section~\ref{sec:montecarlo},
studying  the average number of
crossing of the ``detector  surface'' (at the altitude
$h = 380$~Km) for all  albedo particles   that contribute to the flux
at high altitude.
In the  region $|\lmag| < 0.2$ this
 average number is $\langle n_c \rangle  \sim 20$ for 
$E_k > 0.3$~GeV,  and  $\langle n_c\rangle \sim 14$ for $E_k > 1$~GeV. 
The  energy dependence of $\langle n_c\rangle$  reflects the 
faster  longitude drift of  high momentum  particles.
Since each particle contributing to the albedo flux
will  interact a single time in  the atmosphere,  
the quantity   $\langle n_c\rangle^{-1}$ is a good estimate of
the suppression  factor.

This  discussion can  be summarized  as follows.
When  the primary   cosmic rays interact in the atmosphere,
a small  fraction of the   incident energy flux  ``rebounds''
in the form of  outgoing ``albedo particles''.
Charged  particles  below the magnetic  cutoff remain
trapped in the geomagnetic field and    populate the
``second--spectra''   observed  at the space shuttle orbit.
Eventually these charged particles are  reabsorbed in the atmosphere, with
 peculiar  angular and   spatial 
(interaction point)    distributions.
The proton component of the second spectra  produces
neutrinos   with a  qualitatively  estimated
event rate of order 10$^{-2}$~events/(kton~year),
that is  neglibly small  when compared with the  observed rate
($\sim 100$  in the same units).
This  small  rate  can  be qualitatively understood  observing that:
 (i) only $\sim 1\%$ of all showers  produce  an  albedo proton,
(ii) the   fraction of the  incident c.r. energy  flux
transformed into albedo  protons   is
of order   $\sim 10^{-3}$,  (iii) the  $\nu$  event  yield of 
low  energy  $p$  is  suppressed by kinematical effects.

\section{Conclusions}
In this  work we have analysed the origin of 
the sub--cutoff spectra  of cosmic  rays
 measured  at an altitude of $\sim 400$~km  by the AMS  detector.
The natural  source  mechanism for  these  fluxes
is the  production of secondary   particles in the atmosphere, injected
into  trajectories that  reach  high altitude as  ``albedo particles''.
This simplest
hypothesis allows  to  naturally
 explain   the  remarkable  qualitative
features of   these subcutoff fluxes, such as the high $^3$He/$^4$He  and
$e^+/e^-$ ratios, the  separation into  short and long
lived  particles,  and the restrictions in the extension  of the source
and  sink regions.
We have performed a straightforward   calculation
of the $p$ flux, that  requires less  computer  power  than
the calculation of  the $e^\pm$ fluxes.
In the calcuation we  have used a   rather crude model  for the
production of   nucleons  in the backward hemisphere 
of  a c.r.   interactions. Notwithstanding this  limitation we  obtain
a  reasonably good agreement  with  the  AMS  data,
confirming the results of Derome et al \cite{derome}.

The  long time  of  flight of many  particles in  the sub--cutoff
fluxes, is a  consequence of  the structure of the geomagnetic  field.
Because of the shape of  the $L$  magnetic  shells,
 or in less sophisticated language, because of the offset
of the dominant dipolar component of the   field with respect to the
Earth center,  the trajectory of  the trapped particles   generated 
close to the surface can remain for  a long  time at high altitude,
with a   guiding  center,
that oscillates in  latitude, and    drifts  in longitude
for  as  much as  close to one  complete   revolution around the Earth.
Also  all the  other  ``striking''  properties  of the
sub--cutoff particles  can be given  simple, qualitative explanations.  

The correct  description of  the cosmic  ray  fluxes
reaching then Earth  is important for the interpretation
of the atmospheric  neutrino  data, that are giving evidence
for  the existence of new physics beyond the standard  model.
In this respect the existence if the subcutoff  fluxes
appears to  be potentially important, especially  because
they  exhibit  striking  patterns in the spatial   distributions
of their  points of  origin  and absorption.  
A quantitative  analysis  however reveals that the contribution
to the neutrino event rates of  nucleon  parents
that  enter the $p$ ``second spectrum'', is   a fraction
below 0.1\% of  the observed rate, and is therefore negligible.

On the other hand the data of AMS,
allow  to test experimentally, some  important elements
in the calculation of the  atmospheric  neutrino  fluxes:
 (i)  the hypothesis  of the
isotropy of interplanetary space cosmic  rays
 for  the large angular scales relevant for
atmospheric neutrino  studies,
and (ii)  the  reliability of   the  modeling of geomagnetic  effects
to  determine which trajectories are allowed and
which ones  are forbidden.
No  significant deviations from  the expectations   have been detected,
setting the best  direct experimental  constraints on these   issues.
Note that the study of the second spectra fluxes, 
are  an excellent test of  the quality of our  description of  geomagnetic
effects, since to  obtain agreement with the data 
one needs  an accurate description of  the field and   of particle 
propagation in the field.

A detailed  understanding  of the fluxes of  charged particles  in  
near Earth orbit is important also as a input to the calculation 
of backgrounds  for  scientific instruments aboard satellites
like  for  example the planned  AGILE and GLAST 
high energy  $\gamma$--astronomy telescopes.
A work  on this  topic is in preparation \cite{plan}.

It is interesting  to consider the relation between 
the fluxes of  sub--cutoff particles  observed  at the
altitude of the  space shuttle orbit and the 
fluxes of  trapped  particles in the radiation belts.
In our  view,  a conceptual  distinction between the 
two  populations,    can be   tentatively made as  follows: in one case 
(``second spectra'' fluxes)    the position of
origin   of  a  particle  
(and also its  expected  position of absorption when  the  particle
is detected  non destructively)    can  be    estimated  rather
reliably   and is   close to the surface of the Earth  where the
atmosphere  is  dense;
while in the second case (radiation belts particles)
the   trajectory  treated as a classical 
trajectory in a static  magnetic  field  has  no ``origin''
and no  ``end''  if energy loss mechanisms  are neglected.
The second case  is  possible because in a quasi--dipolar magnetic  field 
there  are many  trapped trajectories
that  remain for an  infinite amount of  time (both  in  the past and
in the future)  in the space 
bound     by   two  finite  radii $r_{\rm min}$ and $r_{\rm max}$.

In the  first  case the injection of the particles
in the population  is  ``direct'',     in the sense that  they are
generated   either at the primary particle 
interaction point or ``not too far'', 
in the  subsequent shower.
The  ``classical''  definition of   the ``cosmic ray albedo''
correspond closely  to this concept  of  a population of
secondary   particles produced in cosmic  rays   showers,
that  can  reach high altitude, even if perhaps 
not all  properties  of the albedo particles
have been   clearly  understood.
For  example the  distinction 
commonly made  in the literature \cite{albedo} 
 between  an upgoing ``splash  albedo''    and a   down--going ``reentrant
albedo'',   has  sometimes  been interpreted as  having   the implication
 that  ``spash albedo'' particles are  generated 
in the vicinity of (below) 
the detector,  and  ``reentrant albedo'' particles are  produced
at  the point of  opposite (magnetic)  latitude and 
 approximately same  longitude.
It is indeed  true that  most albedo particles  
are never  ``reflected'' by the  magnetic  field,  and  are reabsorbed
at approximately the same longitude,  however
also in  the light of the AMS  results   one must understand that
albedo particles can ``bounce''  also  many times  between mirror  points
in the northern  and  southern hemisphere, with  a  long drift
in longitude\footnote{I'm grateful to Yu.Galaktionov  for
discussions on this  point.}.

The mechanism   considered  as the main  source of the charged particles
trapped  in the radiation  belts, has  been
the   decay of  secondary neutrons \cite{Walt}.
Since the $n$ lifetime  is long, the  $n$ decay  point
(that is  the creation point of a  particle in the belt)
is  only   weakly correlated 
with   a  c.r. shower.

As a final comment, I would like to  speculate, that 
a  significant  source of
particles  in the inner belt    could be 
related to the existence of  populations of
long lived   albedo  particles ($p$   and $e^\pm$) 
 produced at intermediate magnetic latitudes. 
These particles  have  a
confinement volume with    similar  shape  to the inner belt 
(see for example  figure~\ref{fig:traj1} and~\ref{fig:traj2}),
 and can be the source
of  ``permanently''   confined  lower   rigidity  particles
via their interactions  with  the residual  atmosphere 
at very high altitude.

\vspace {0.4 cm}
\noindent {\bf Acknowledgments}  I have to express my gratitude
to
J.~Arafune, 
T.Gaisser,
M.Honda,
T.Kajita,
J.~Nishimura
and S.Vernetto
 for  useful  discussions.
I'm also  grateful  to the 
R.~Battiston, Yu.~Galaktionov, E.~Fiandrini, B.~Bertucci and  G.~Lamanna
for  discussions  about the AMS data.
Finally my thanks  to S.~Ting for an invitation to
the   AMS  workshop in the Ettore Majorana Center in  Erice.

\clearpage

\appendix

\section*{Appendix: Geomagnetic effects}
\setcounter{section}{1}
\label{sec:geomagnetic}

\subsection{The Earth's  magnetic  field}
\label{subsec:earth}
It is  well known  that the  Earth's magnetic   field
to a first approximation can be described  as  a dipole  field.
In spherical  coordinates\footnote{We  have  chosen the 
 origin of the coordinates  at  the 
dipole  center  and the polar axis  opposite
to $\vec{M}$,  since for the
Earth's  the magnetic  moment  $\vec{M}$  points  south.}
the components of the  field are:
\begin{equation}
B_r = -{ M \over r^3} ~ 2 \, \sin \lambda; ~~~~~~~~~
B_\lambda = { M \over r^3} ~ \cos \lambda.
\end{equation} 
where $M$ is  the magnetic  dipole  moment  and $\lambda$ is  the
magnetic  latitude.
For the Earth
$M \simeq 8.1 \times  10^{25}$~Gauss~cm$^3$,   that   corresponds  to
an equatorial magnetic  field   at  the surface.
$B_{\rm eq} \simeq M/R_\oplus^3 \simeq 0.31$~Gauss.
The  field  lines    for  a dipolar  field have the
 form:
\begin{equation}
r = r_0 ~\cos^2 \lambda
\label{eq:field-line}
\end{equation}
The    module   of the field  $|\vec{B}|$
along each field  line 
has    its   mimimum  value  on the  equatorial plane
($\lambda = 0$)   at the point with the largest
distance from dipole  center (with $r = r_0$).

It is  well  known   that   
the geomagnetic field  
is  significantly  different 
from the exact  dipolar  form.
These  deviations   of the field
from the dipolar form are essential  for  an  understanding of
 the  properties  of the sub--cutoff  fluxes.
The sources  of the     magnetic  field   can  be very naturally
divided
into  ``internal sources''
(electric  currents  inside  the Earth), and 
``external  sources''   (electric currents 
in  space).    
The   contribution to the field of the external sources
exhibits  variations  also with  very   short   (hours) time scale,
connected with  the position and magnetic activity  of the sun,
while the  contribution of the ``internal  sources'' 
 varies only  on   much longer time  scales,
with a secular drifts of the magnetic  poles.
The    magnetic    field  due  to the external sources
is the dominant  contribution    to  $B$ 
at a  distance of several Earth's  radii,
but  represent  only  a   small  perturbation  in the  vicinity of the
Earth  and will be neglected  in this  work,  where   the geomagnetic field
will be described  as  in the
International Geomagnetic Reference Field (IGRF) model \cite{IGRF}
that  is an  empirical representation
based  on a  multipole expansion.
The coefficients 
of the different multipole terms 
(often called  the Gauss coefficients)
are  slowly time  dependent.
In the  numerical  work we  have used the IGRF field  that  corresponds
to the 1st of january  2000.

It has   been  known   for  a  long time, that  if one  wants  to
describe the   geomagnetic  field  with a simple  dipole,  one
obtains  a  significantly  better  fit    with a dipole  that is not
only
``rotated''  with   respect  to the Earth axis, but it is also
``offset''  that is    it  has  an origin  that does not  coincide
with the Earth's center.
In the (standard) expansion of the field   used  in the IGRF  model   the
origin of all 
multipole   terms   is the Earth's center, however it is possible to  
approximately ``reabsorb'' the quadrupole contributions
redefining the dipole   moment and     the position of its  center.
There  is  no  unique  well defined way to   perform  this   redefinition
of the dipole and different  algorithms  have been   used 
for  different  applications.  Qualitatively  however   the effect
is    quite  clear.
The  dipole  offset (planet center  to  dipole  center)
for  the  Earth is of order  $\sim 450$~km   and  a vector 
from the  Earth's  center to the dipole  center has  a latitude 
 $\sim 18^\circ$   and  a longitude  $\varphi_{\rm dipole} \sim 140^\circ$.
The  approximate offset of the dipole axis
with respect to  the Earth's  center  is of crucial
importance  for  the understanding of  the second spectra
observed  by AMS  and is illustrated in  
fig.~\ref{fig:equator2}.

\subsection{Motion of charged  particles in a magnetic  field}
\label{subsec:motion}
The properties of trajectories of   charged  particles  in a
magnetic  field  are  a  classic  subject with a   rich literature
(see for example \cite{Jackson}
for  an  elementary introduction, or
\cite{Rossi-Olbert}  for a  more detailed  discussion).
Here we  will only  recall  some
simple   results   that  will  be  used   in this work.

In a static homogeneous    magnetic field  the motion of 
a charged particle  is a  helix, 
with  gyroradius  $a$:
\begin{equation}
a({\rm Km}) = 33.3 ~ {p_\perp({\rm GeV}) \over B({\rm Gauss}) } 
\label{eq:gyro}
\end{equation}
This  motion  can  be analysed   as the  combination  of
 a   rotation in a
plane  orthogonal to the field  lines,   accompanied by   a uniform  motion
along the field  lines.

In  a non uniform static  magnetic field
where   the  distance scale $L$ 
of the  field   variation  $L \sim | B^{-1} \partial B/\partial x_j |^{-1}$
is much  larger  than the  gyroradius $a$ ($L \gg a$),
the   motion  of a  charged  particle
can again be   decomposed as    the  rotation 
in a plane  orthogonal to the field  lines around  a   point (the 
``guiding  center'')   that   has  a motion
both  along  and  across    the  field  lines.
The   motion parallel  to the field  lines is  controlled  by the 
variation of the field  intensity  along the  field  line.
This  behaviour  can  be
deduced   from the (adiabatic)   conservation
of the magnetic  flux  ($\pi a^2 B$)   through a particle's 
circular  orbit.
The  conservation of the magnetic  flux  can be written in the form:
\begin{equation}
v_\perp^2 = v_{\perp 0}^2 \; {B \over B_0} 
\label{eq:bounce}
\end{equation}
Using   the fact that $v_\parallel^2 + v_\perp^2$ is  constant,
one  can deduce the equation
\begin{equation}
{\partial v_\parallel \over \partial \ell} \simeq  -{v_{\perp 0}^2
\over 2\, B_0 }~ {\partial B (\ell) \over \partial \ell}
\end{equation} 
($\ell$ is  the  distance  along the field   line)  that describe the 
motion parallel  to the field line.
The increase  of  $B$  along the 
field   line  has a ``repulsive effect''   and is  at  the 
basis  of the ``magnetic  mirror'' effect. 

When the    gradient of the field  has a non vanishing
component    $\nabla_\perp B \ne 0$, or when  the 
field  lines  are  curved,    the guiding  center
has  also  a  ``drift'' motion   orthogonal  to the  field.
Positively  and  negatively charged  particles  drift  in opposite
directions.

\subsection{``Allowed''and ``Forbidden'' Trajectories}
\label{subsec:allowed}
The fluxes  of cosmic rays  observed  
at points with  different  magnetic  latitudes,
are dramatically  different, with
the flux  measured  close to the magnetic  equator  strongly suppressed
with respect to the flux  measured at high  magnetic  latitudes.
The  discovery  of the   ``latitude effect'' \cite{Clay,Compton}, lead
to the  understanding that the ``cosmic radiation''  was  mostly composed
of  charged particles.
 Soon Bruno Rossi \cite{Rossi}   observed  that 
the geomagnetic  effects should produce  an  east--west  asymmetry,
whose sign  would determine  if   most c.r.  are
positively  charged   (excess of east--going particles)  or 
negatively charged (excess of west--going particles).
The  effect was soon detected \cite{johnson,alvarez}, 
determining that most c.r.  have  positive  electric charge.

The latitude and  east--west effects
are the simple consequence of the fact that 
low   rigidity particles  from outer space
cannot  reach the  Earth's surface because of the
geomagnetic  field.
Let us  consider a detector    located at the  position
$\vec{x}$  that measures  a particle of   electric charge $Z$ and
 momentum $\vec{p}$.   To  a very  good  approximation  the past trajectory 
of the detected particle  can be   determined 
integrating the  classical  equations  of motion for a charged particle 
 in an   electromagnetic  field  in the region  around the 
detector (and the Earth).  Reconstructing this 
 past trajectory there  are  three
possible  results: 
\begin{itemize}
\item[(a)]  the trajectory originates from the Earth's surface
(or deep in the atmosphere);

\item[(b)] the  trajectory  remains  confined  in the 
    volume  $ R_\oplus  < r < \infty$   without 
   ever reaching ``infinity'' (where $R_\oplus \simeq 6371.2$~Km is  the 
Earth's  radius);

\item[(c)]   the particle   in the past   was at  very large distances 
     from  the Earth. 
\end{itemize}
Trajectories belonging to the classes (a) and (b)  are considered 
as ``forbidden'',  because no primary cosmic  ray  particle
can reach  the Earth   from  a large distance
 traveling along one of these trajectories.
All other  trajectories are allowed.

If we  consider a  fixed detection position 
$\vec{x}$  and  a fixed direction 
$\hat{n}$,     to a  reasonably good approximation 
the  trajectories     of all positively charged particles  with  rigidity
larger  (smaller)  than  a  cutoff 
$R_+^{\rm cutoff} (\vec{x}, \hat{n})$    are  allowed (forbidden).
This  is exactly true for a dipolar field  that  fills the  entire
space.  In this case the solution  (the ``St\"ormer cutoff'')
can be written  down as an analytic expression \cite{Stormer,pl-3d}.
In the more  general case    it is necessary to
study the problem numerically.

The effect of the geomagnetic  field on an   isotropic  
interplanetary flux is simply to ``remove'' the  particles form
the forbidden trajectories, without  deforming the shape
of the spectrum.
This  can  be  deduced from the Liouville  theorem,
with the assumption  that the field  is static \cite{Lemaitre-Vallarta}.

\subsection{Trapped particle trajectories  in a dipole  field}
\label{subsec:trapped}
The motion of  trapped  particles in  a dipolar magnetic   field
can be  simply  understood as  the combination  of 
three periodic  motions
having    three  very different  characteristic  frequencies.
The  first motion  is  simply the  gyration   around the 
magnetic    field lines
with a  frequency
\begin{equation}
\Omega_{\rm  gyro} =  {Z e B \over E} \simeq 9009
~ {Z \, B({\rm Gauss}) \over
E ({\rm GeV}) }~ {\rm sec}^{-1}
\label{eq:omega-gyro}
\end{equation}
The second  component  of the 
 motion  is  a   constant  velocity  longitude  drift.
Positive  particle drift  westward  (toward  decreasing  longitude).
Negative  particles  drift eastward  (toward  increasing longitude).
The  drift is   simplest  for  particles   that move
in the equatorial plane.
In this  case it is elementary to show  that   the guiding  center 
moves uniformly on a circle   centered on the dipole  with a
frequency:
\begin{equation}
\Omega_{\rm drift} = {3 \over 2} ~ {r \over M} ~ \left ( { p \, \beta
\over Z e} \right )  \simeq  1.19~ {r \over R_\oplus} ~ 
{ p({\rm GeV}) \,  \beta \over Z}~  {\rm sec}^{-1}
\label{eq:omega-drift}
\end{equation}
(for the numerical  estimate we have used the Earth's dipole moment).
Note  that     the  frequency 
is  $\propto p$, and therefore  the  period to 
perform an  orbit   around  the  Earth   is proportional
to $p^{-1}$.
This   behaviour  is  easily understood  qualitatively 
since the longitude drift  is produced  by the 
dishomogeneity of the  field and the variations of the   gyroradius  of
the particle   as the particles moves.
Higher  momentum particles, with a larger  gyroradius, are
more sensitive  to  the   gradient. 
Note also  the curious  result   that the drift  frequency is 
linear in $r$ that  can also be understood considering the $r$ dependence
of the field gradient.

The third component  of the motion 
is  an oscillation around  the equatorial plane.
A particle  that  finds  itself  on the 
equatorial plane    with a  non vanishing component  of  the momentum
parallel  to the  field  
(that  is with   $\cos \alpha_0 =0$, where
 $\alpha_0$   is the pitch  angle  on the equatorial  plane) 
will  bounce   back and    forth  between (symmetric)  maximum
and  minimum  latitudes. The value of the field  at the bouncing
points  is  given by equation (\ref{eq:bounce})
solving for $v_\parallel = 0$. It is  clear  that the amplitude
of the oscillations  is  determined  only by the  pitch  angle
$\alpha_0$, and is  independent   from the particle  momentum
and electric  charge.
This   latitude  oscillation is to a good approximation an harmonic  motion.
For  small amplitude oscillations   (that is when the pitch angle
in the  equator plane $\alpha_0$ is close to $90^\circ$)
the  frequency  is  amplitude independent:
\begin{equation}
\Omega_{\rm bounce}^0 =  {3 \over \sqrt{2} } { \beta  \over r}  \simeq 
99.8~ \beta~{ R_\oplus \over r_0} ~{\rm sec}^{-1}
\label{eq:omega-bounce}
\end{equation}
When  the amplitude  of the   bounce   increases (with 
 growing  $|\cos \alpha_0|$)  the oscillation
frequency  depends    weakly on  the    amplitude:
$\Omega_{\rm bounce}  (\cos\alpha)  =  
\Omega_{\rm bounce}^0 / \tau_b$ 
with  $\tau_b$ a  dimensionless quantity  that 
is unity  for   $\cos \alpha_{0} = 0$    and  grows  monotonically to
$\tau_b \simeq 1.87$ for  $\cos \alpha_{0} \to 1$.

\subsection{``Bouncing motion''}
\label{subsec:bouncing}
It can   be useful  for the discussion to consider more  closely 
the  ``bouncing''  motion of  the trapped particles, that is the oscillations
of the  guiding center of the particle trajectory
between  two mirror points 
placed  symmetrically in the north and  south  hemisphere.
It is  obvious   that  only oscillations
with  a sufficiently small amplitude are possible,
because  following a  field  line, from a  point $E$  in the
equatorial plane (in either  hemisphere)
the  radius  $r$  decreases, and therefore 
if the amplitude of  the oscillations  are too large 
a particle ``hits'' the surface of the Earth  and  is absorbed.

The condition on the  maximum amplitude can  be translated  in a 
(rigidity independent) condition
 on the   pitch  angle $\alpha_0$  of the  trapped particles
when they are on  the  equatorial plane.
 Let us consider the magnetic  field  line  that  passes  through the 
point $E$ on  magnetic  equatorial plane.
This line   exits from the surface of the Earth  in a point $G_1$
in the southern hemisphere, 
and reenters the Earth's surface  at a point $G_2$
(for a  centered  dipole the points  $G_1$ and $G_2$  have  the same 
magnetic  longitude  and
symmetric   latitudes: 
$\cos \lambda_{1,2} =  \mp \sqrt{R_\oplus /r_0}$).
The value  of  $|\vec{B}|$    along the  field  line
has its   mimimum  at the point  $E$ in the  equatorial plane 
 and grows  monotonically   with
the distance   from $E$   (symmetrically for a dipolar field).
The increase  of the field  along the  field  line  acts  as 
``magnetic mirror'' or repulsion (see equation
(\ref{eq:bounce})).
A  charged  particle  at the point $E$   with pitch angle  $\alpha_0$  
that is  not exactly $90^\circ$,  will have  a  component of  momentum 
parallel to the field  $p_\parallel = p\,\cos\alpha_0$ and will start  moving
along the field line, but the  gradient of the  field 
along the line will reduce  and finally invert
the parallel  component.  The inversion point is the  ``mirror'' point.
 The  component 
of the momentum  parallel   to the field at a point
$P$  along the line    ($\propto \cos \alpha$)
depends  of the value of the field at that particular point.
From  equation (\ref{eq:bounce})  one obtains:
\begin{equation}
\cos^2 \alpha = 1 - \sin^2 \alpha_0~ {B \over B(E)}.
\end{equation}
The    mirror points   ($M_1$  and $M_2$)  are  by definitions
the points where the parallel  momentum vanishes, that is:
\begin{equation}
0 = 1 - \sin^2 \alpha_0~ {B(M_{1,2}) \over B(E)}
\label{eq:mir}
\end{equation}
The requirement  that the two mirror points  are above sea level
can then be written as:
\begin{equation}
 |\sin \alpha_0|  \ge 
\left [ {B(E) \over B(G_{1}) } \right ]^{ {1\over 2} },
~~~~~~~
 |\sin \alpha_0|  \ge 
\left [ {B(E) \over B(G_{2}) } \right ]^{ {1\over 2} },
\label{eq:mirror1}
\end{equation}
For a centered dipole field the two conditions 
in (\ref{eq:mirror1})  are of course identical.
Substituting  the  explicit  expressions 
one obtains  the  condition:
\begin{equation}
|\sin \alpha_0| \ge  
r_0^{-{5 \over 4}} \;(4r_0 -3)^{-{1 \over 4}}
\label{eq:mirror0}
\end{equation}
(where $r_0$ is  the radius of the equator point in units of $R_\oplus$).
For example  for   an altitude  of 100~Km equation~(\ref{eq:mirror0})
tells us that  only  particles with  pitch angles between 75$^\circ$
 and  105$^\circ$   can   ``bounce''   without  being  absorbed  by the
Earth.   When  $r$ increases,  larger  amplitudes  of the latitude
oscillations  become  possible.

\subsection {Examples of trajectories in the Earth's field}
\label{subsec:trajectories}
\begin{figure} [t]
\centerline{\psfig{figure=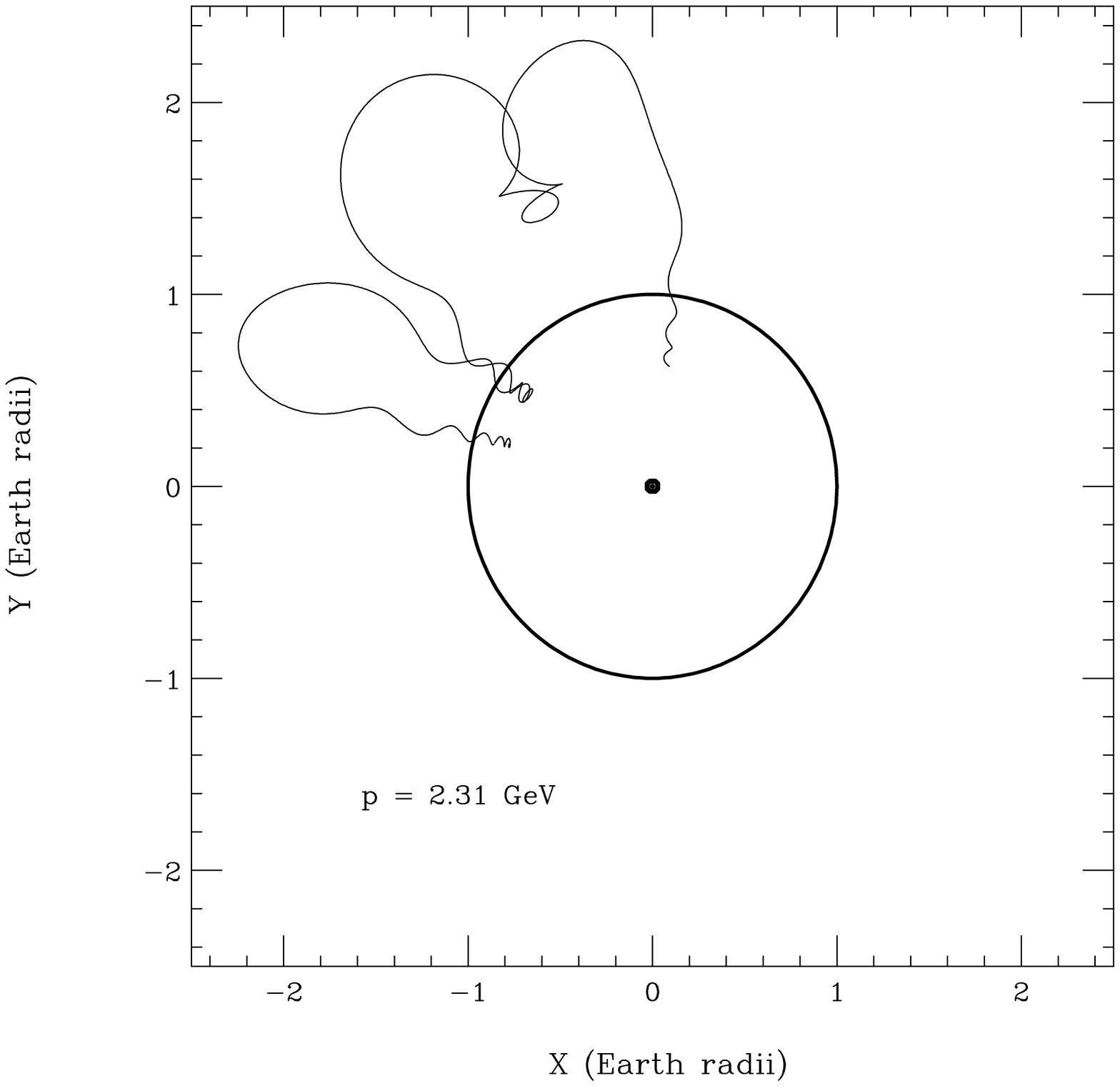,angle=90,height=9.0cm}}
\vspace {0.5 cm}
\centerline{\psfig{figure=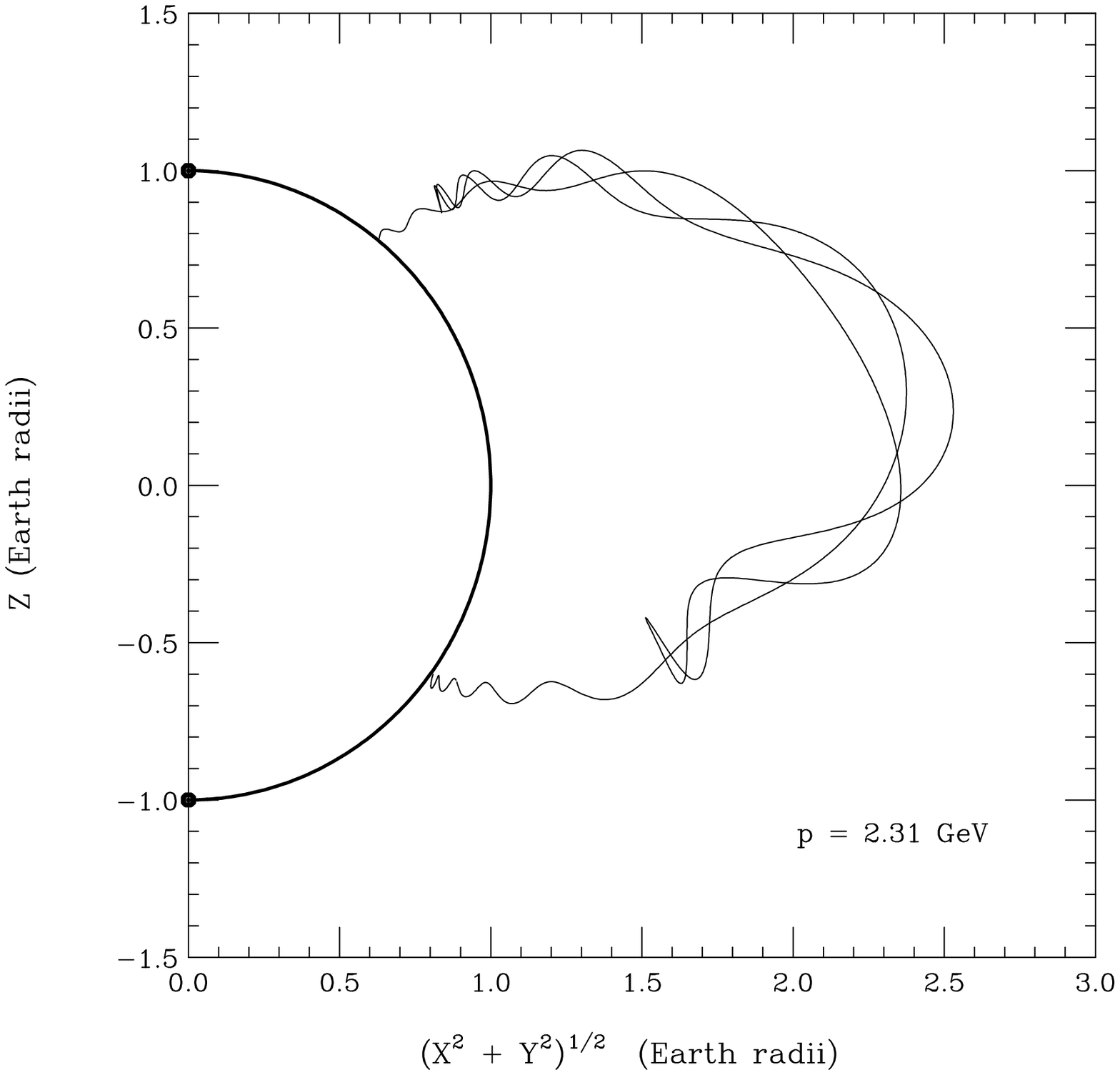,angle=90,height=9.0cm}}
\caption {\small
 Example of the  trajectory
of  a secondary proton  in the geomagnetic field.
Top  panel: projection  of the  trajectory in the Earth's 
equatorial plane.
Bottom panel: projection in the plane $(\sqrt{X^2 + Y^2}, Z)$.
(Geographical  coordinates).
\label{fig:traj1}  }
\end{figure}

\begin{figure} [t]
\centerline{\psfig{figure=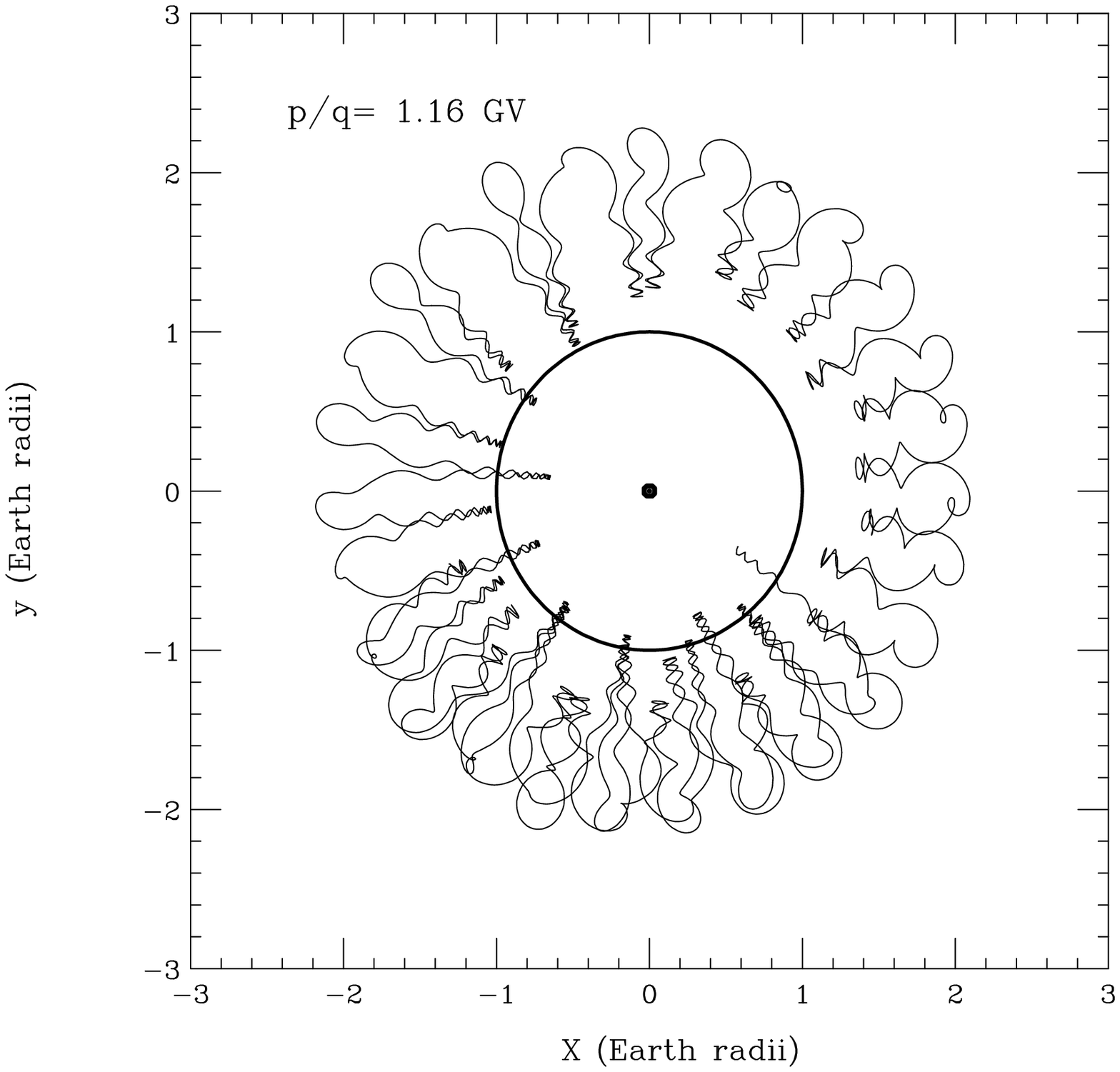,angle=90,height=9.0cm}}
\vspace {0.5 cm}
\centerline{\psfig{figure=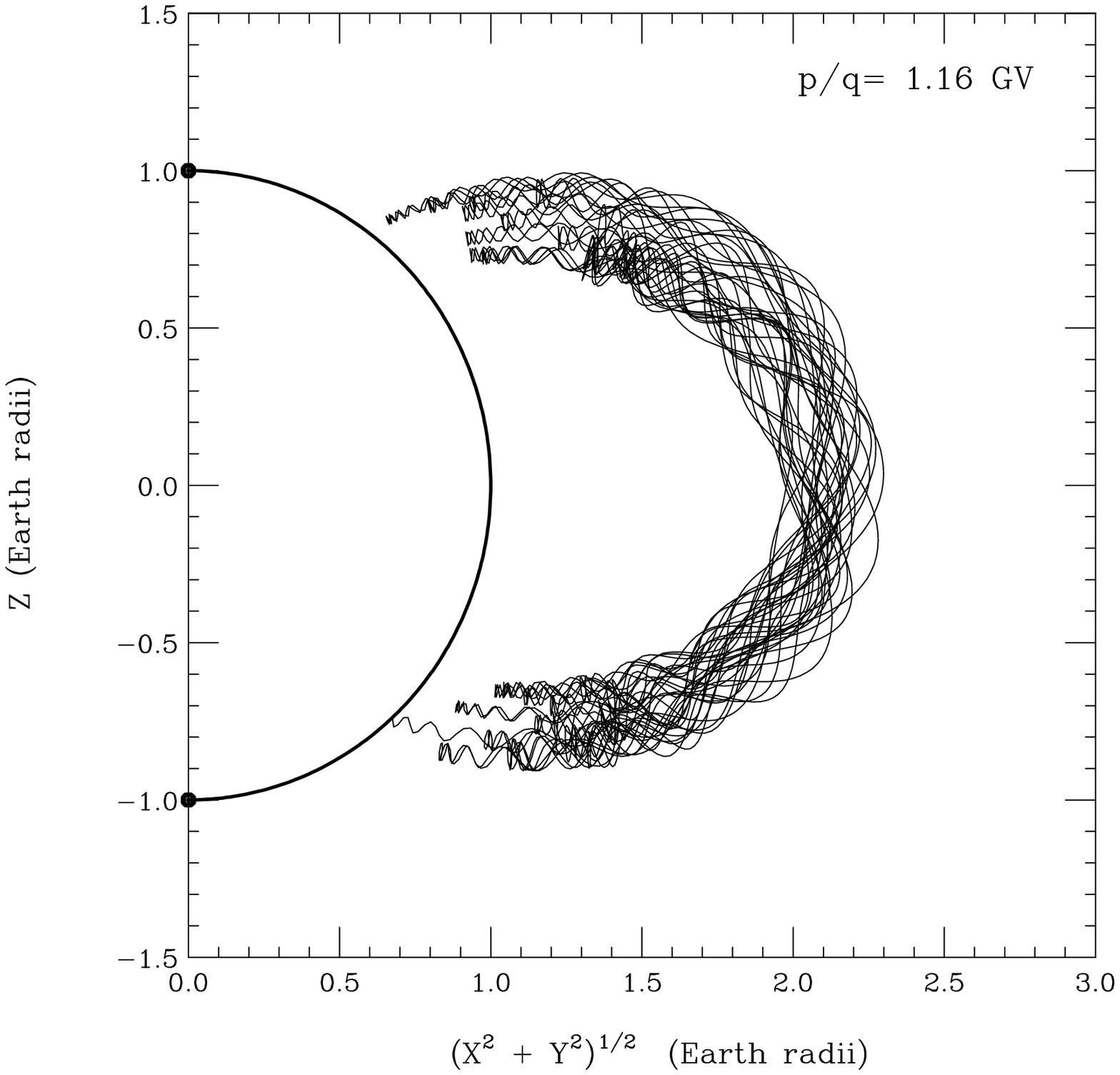,angle=90,height=9.0cm}}
\caption {\small
 Example of the  trajectory
of  a secondary proton  in the geomagnetic field.
This  proton    was traced for $10^6$~km 
without interactions in the atmosphere.
Top  panel: projection  of the  trajectory in the Earth's 
equatorial plane.
Bottom panel: projection in the plane $(\sqrt{X^2 + Y^2}, Z)$.
(Geographical  coordinates).
\label{fig:traj2}  }
\end{figure}

\begin{figure} [t]
\centerline{~~~~~~~~~~~~~\psfig{figure=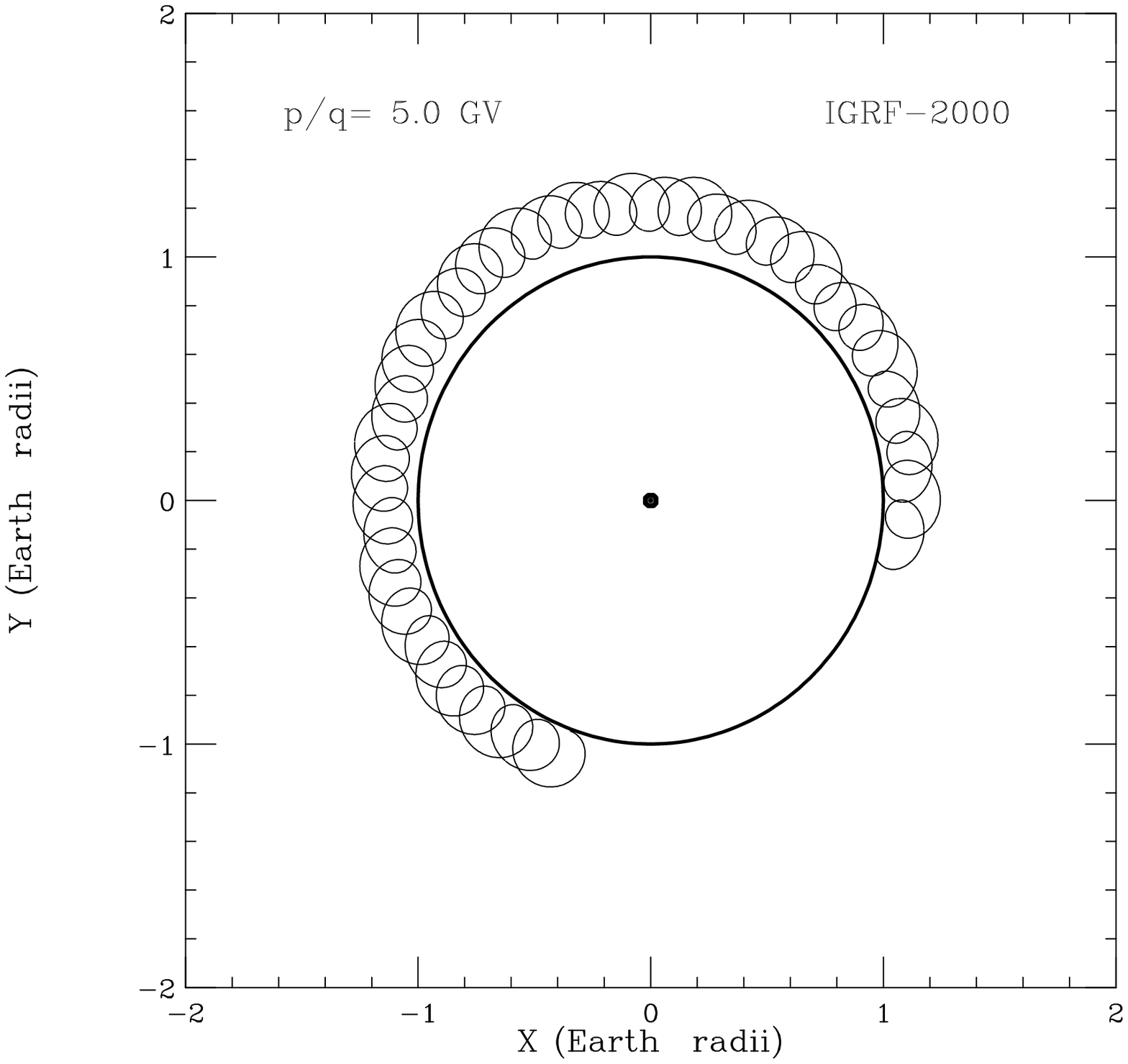,angle=90,height=9.2cm}}
\vspace {0.25 cm}
\centerline{~~~~~~~~~~~~~\psfig{figure=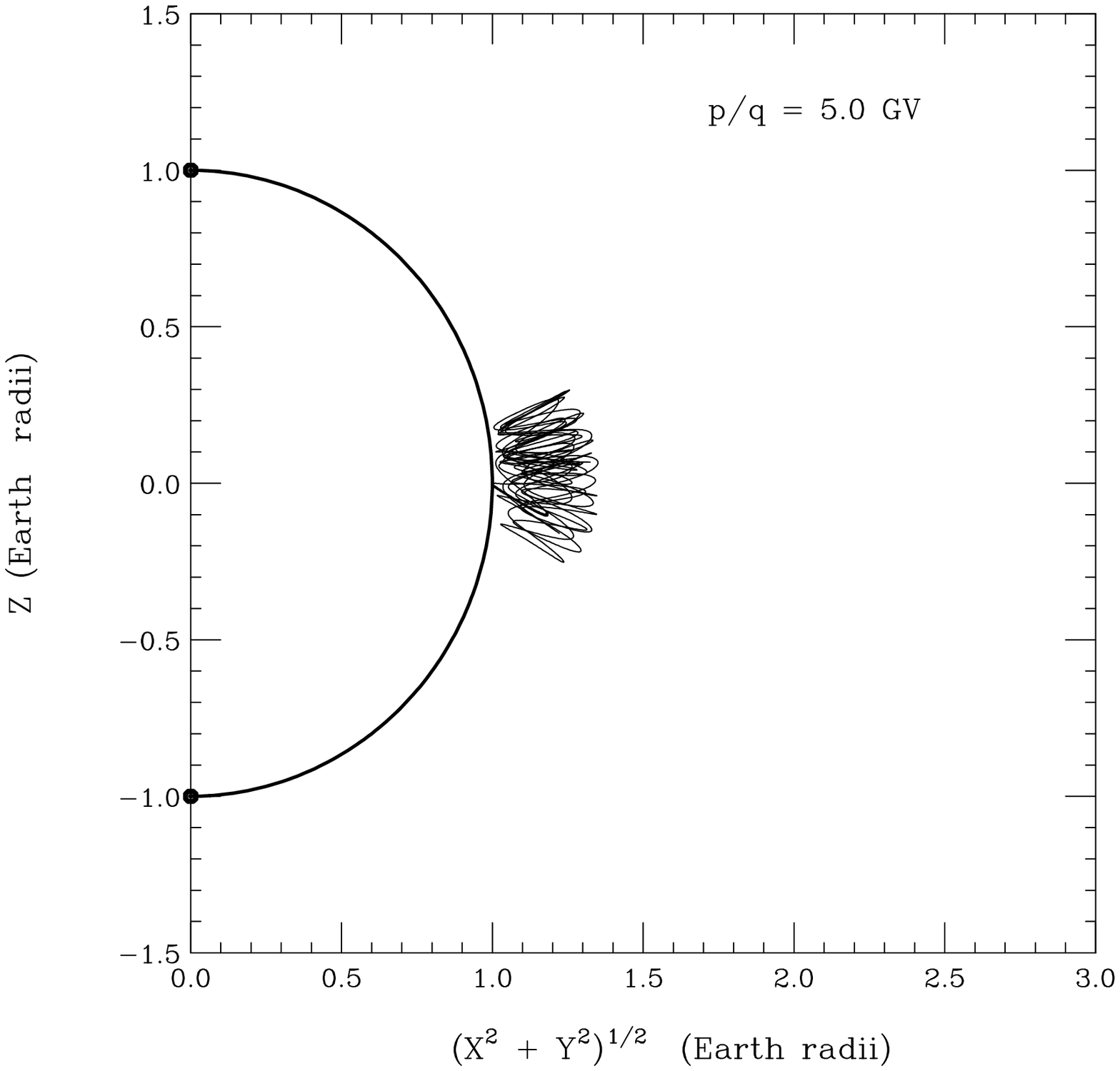,angle=90,height=9.2cm}}
\caption {\small 
 Trajectory in the   geomagnetic field
of  a positively  charged particle with rigidity 5.0~GV. 
The  particle  production point is  near  the equator, at
 an altitude of 50~Km and with 
an initial  momentum  pointing toward geographical east.
Top panel: projection of the trajectory in  the
($X$,$Y$) plane;
bottom  panel: projection  in  the
($\sqrt{X^2 + Y^2}$,$Z$) plane.
\label{fig:ex1}  }
\end{figure}

\begin{figure} [t]
\centerline{\psfig{figure=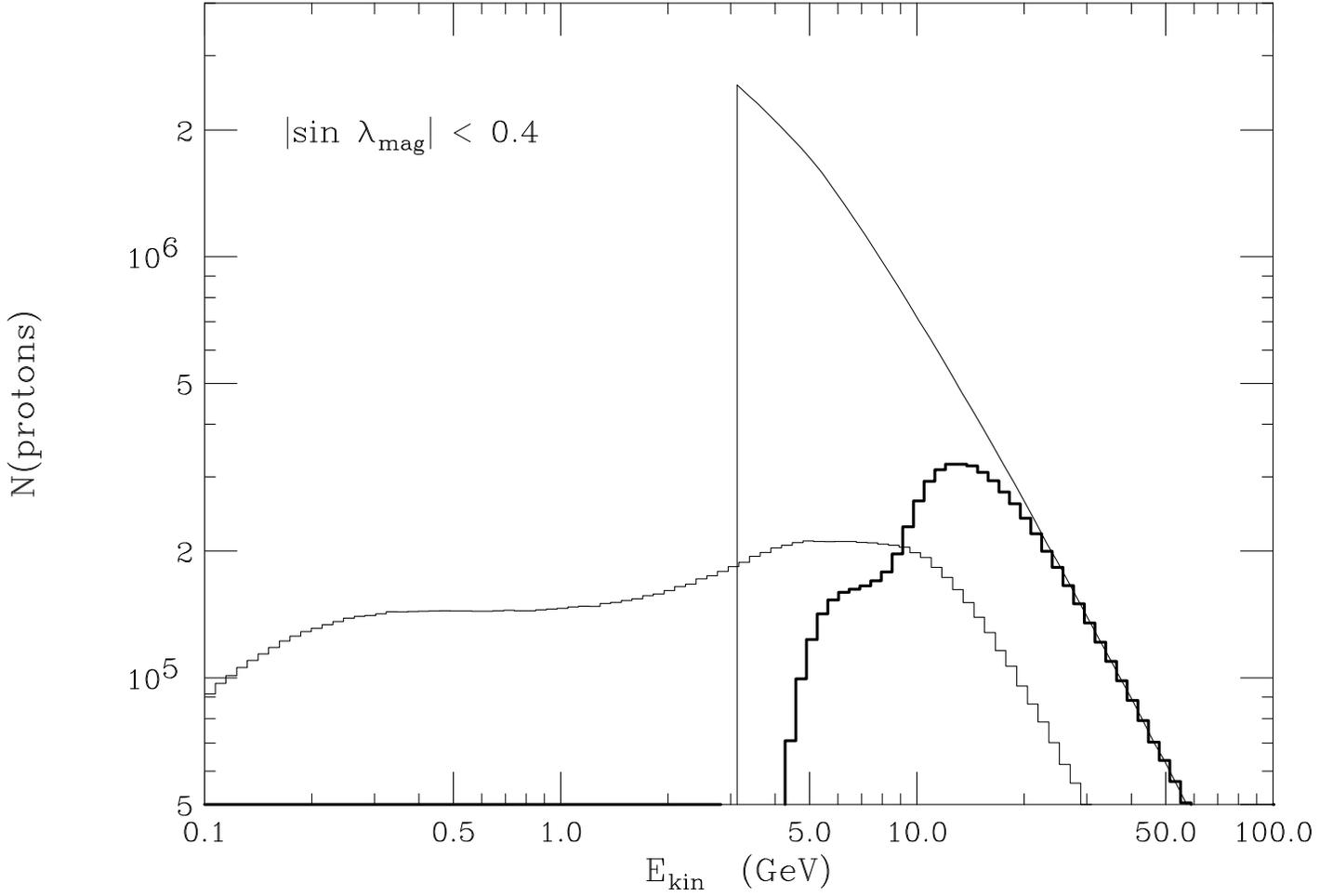,angle=90,height=12.9cm}}
\caption {\small 
 The  thin--solid line  shows 
as a function of kinetic energy per nucleon  the  shape of the interplanetary
nucleon  spectrum  assumed in the Montecarlo  calculation.
The thick histogram shows  (with correct relative normalization)
the  energy  distribution of all nucleons that interact
in the Earth  magnetic  equatorial 
region $|\sin \lmag| < 0.4$. The effects  of
the  geomagnetic  cutoff that  ``forbids'' low rigidity
trajectories are evident.  The  shoulder present at
low energy is  due to nucleons that  reach the Earth in the form
of bound nuclei, and 
that have  higher rigidity for the same energy per nucleon
(since $A/Z \simeq 2$).
The thin  histogram  is the  energy distribution
of all secondary protons  generated in   the
showers of the primary particles.
\label{fig:energy}  }
\end{figure}

\begin{figure} [t]
\centerline{\psfig{figure=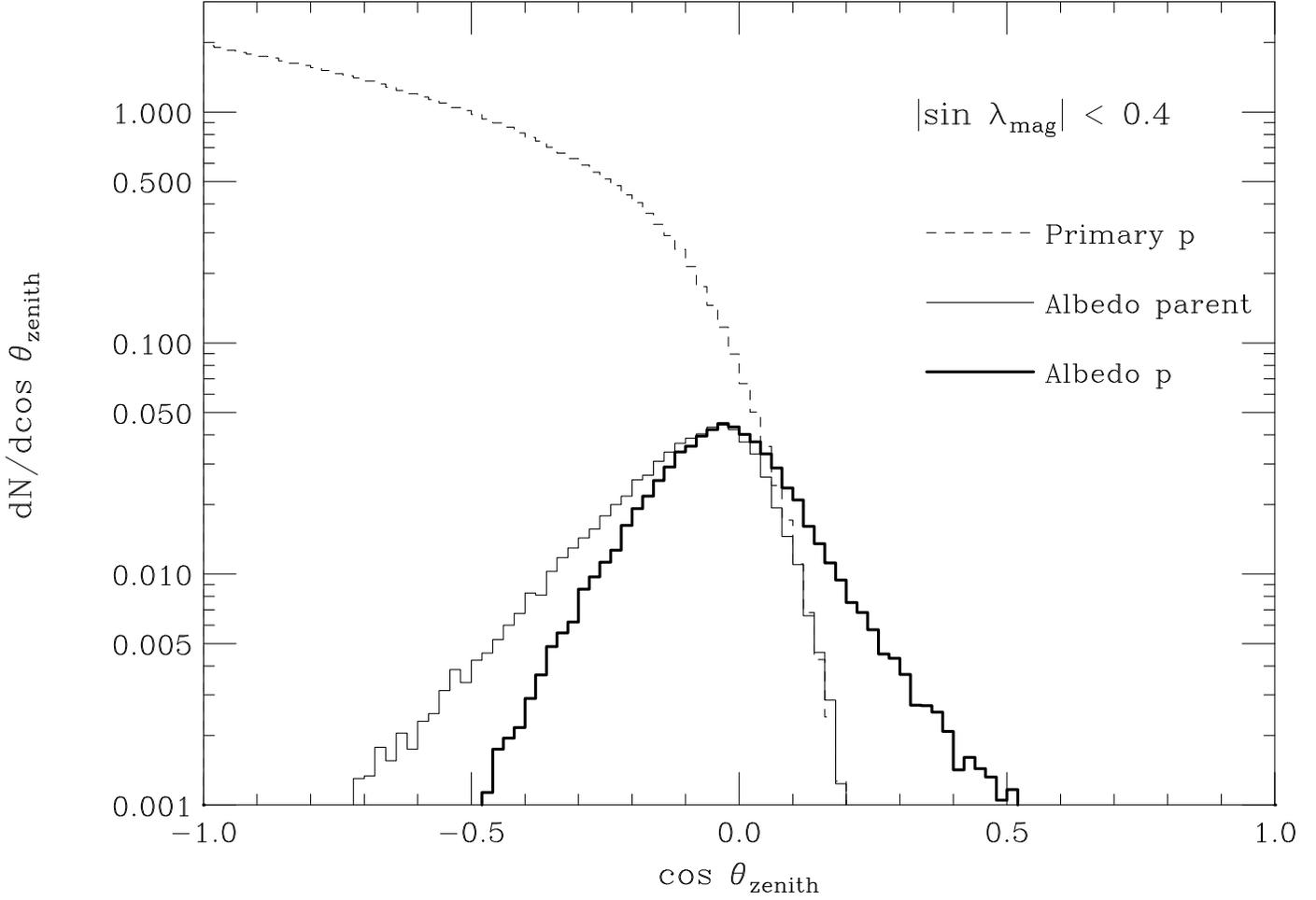,angle=90,height=12.9cm}}
\caption {\small  
Dashed histogram: zenith  angle distribution 
at the interaction point  of all primary cosmic ray nucleons  interacting
in the Earth's  magnetic  equatorial  region 
$|\sin \lmag| \le 0.4$
($\cos \theta = -1$ corresponds to a vertically down--going particle).
The  histogram to a good approximation  has the shape
$\propto |\cos \theta|$,  however  few particles,  bent by the geomagnetic
field interact  having an  ``up--going'' direction.
Thin--solid histogram:  selection  of  nucleons  that
produced  ``albedo protons'', that is 
 a  secondary  $p$   with a trajectory that  reaches 
an altitude larger than 380~Km.
Thick---solid  histogram: 
zenith angle distribution  at the production point of albedo $p$.
\label{fig:zeni}  }
\end{figure}

\begin{figure} [t]
\centerline{\psfig{figure=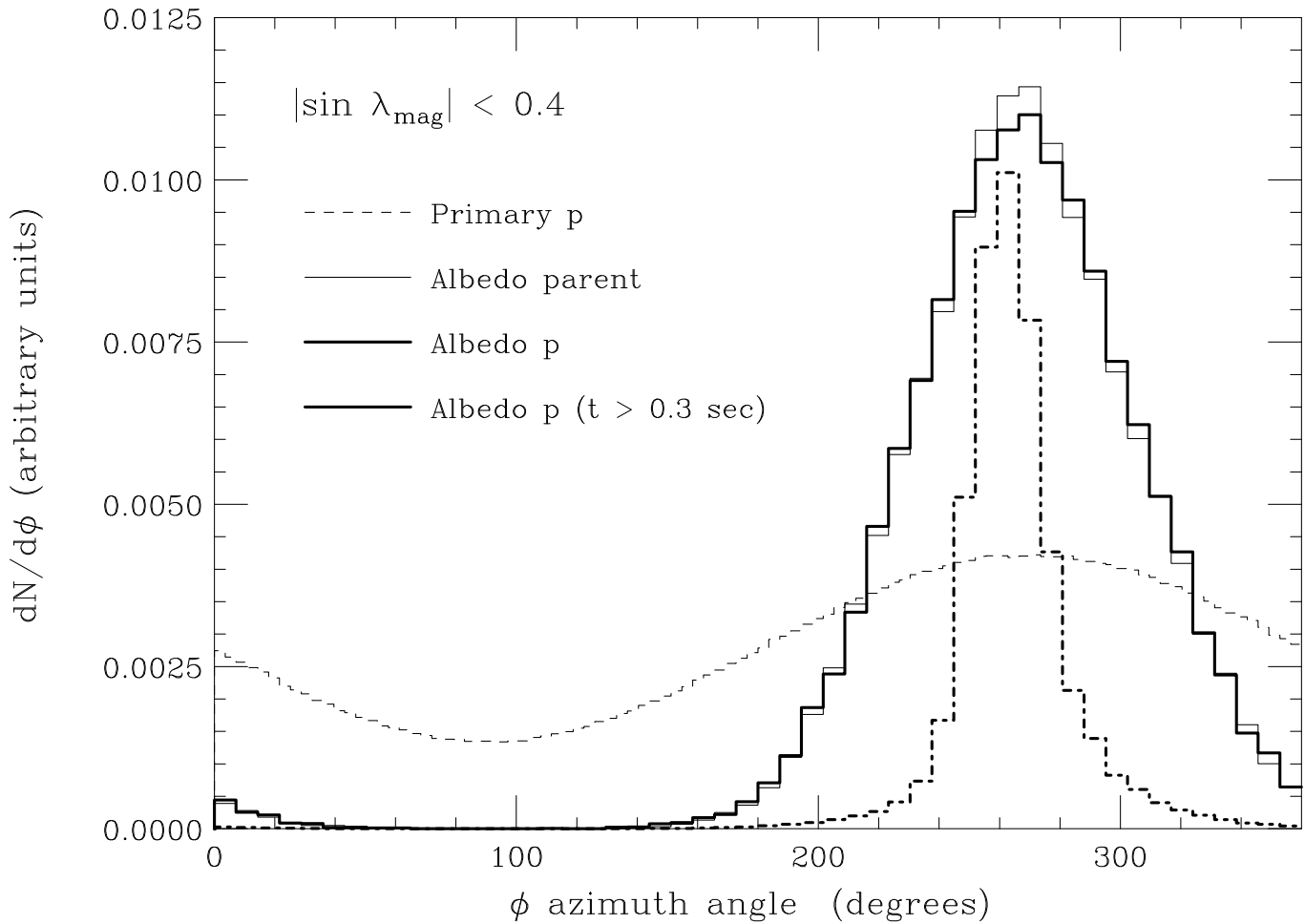,angle=90,height=12.9cm}}
\caption {\small  
Azimuth angle distributions.
Dashed histogram: primary cosmic rays at the interaction point;
thin--solid histogram:  
primary particles  that produce
albedo $p$;
thick--solid histogram, albedo  $p$ at the creation point;
thick--dotdashed histogram: long lived albedo $p$   ($t > 0.3$~sec).
\label{fig:azi}  }
\end{figure}

\begin{figure} [t]
\centerline{\psfig{figure=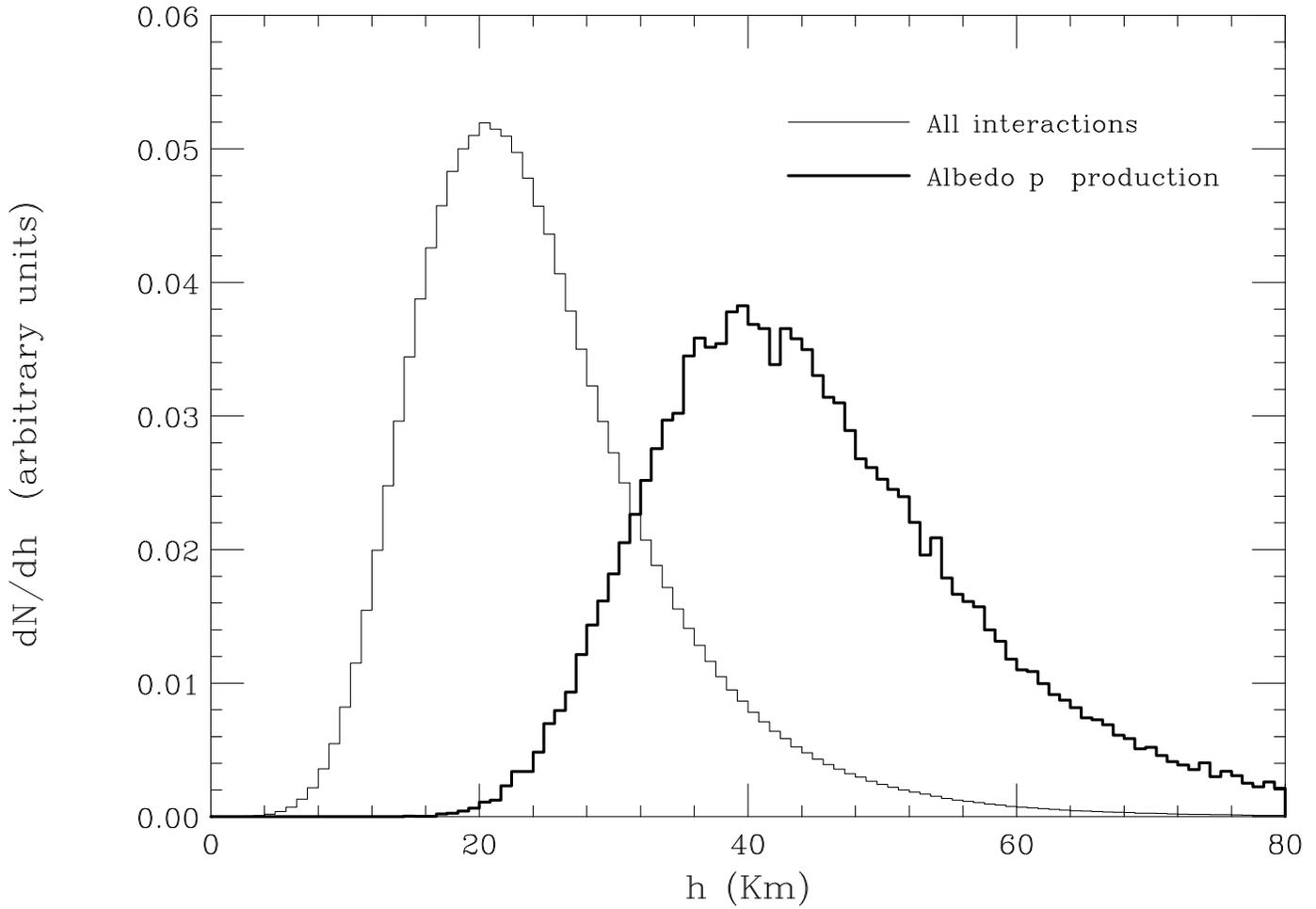,angle=90,height=12.9cm}}
\caption {\small  
Thin  histogram: distribution of the 
altitude of the interaction points   of primary 
cosmic  rays.
Thick histogram: distribution of  production points of 
albedo protons.
\label{fig:hh}  }
\end{figure}

\begin{figure} [t]
\centerline{\psfig{figure=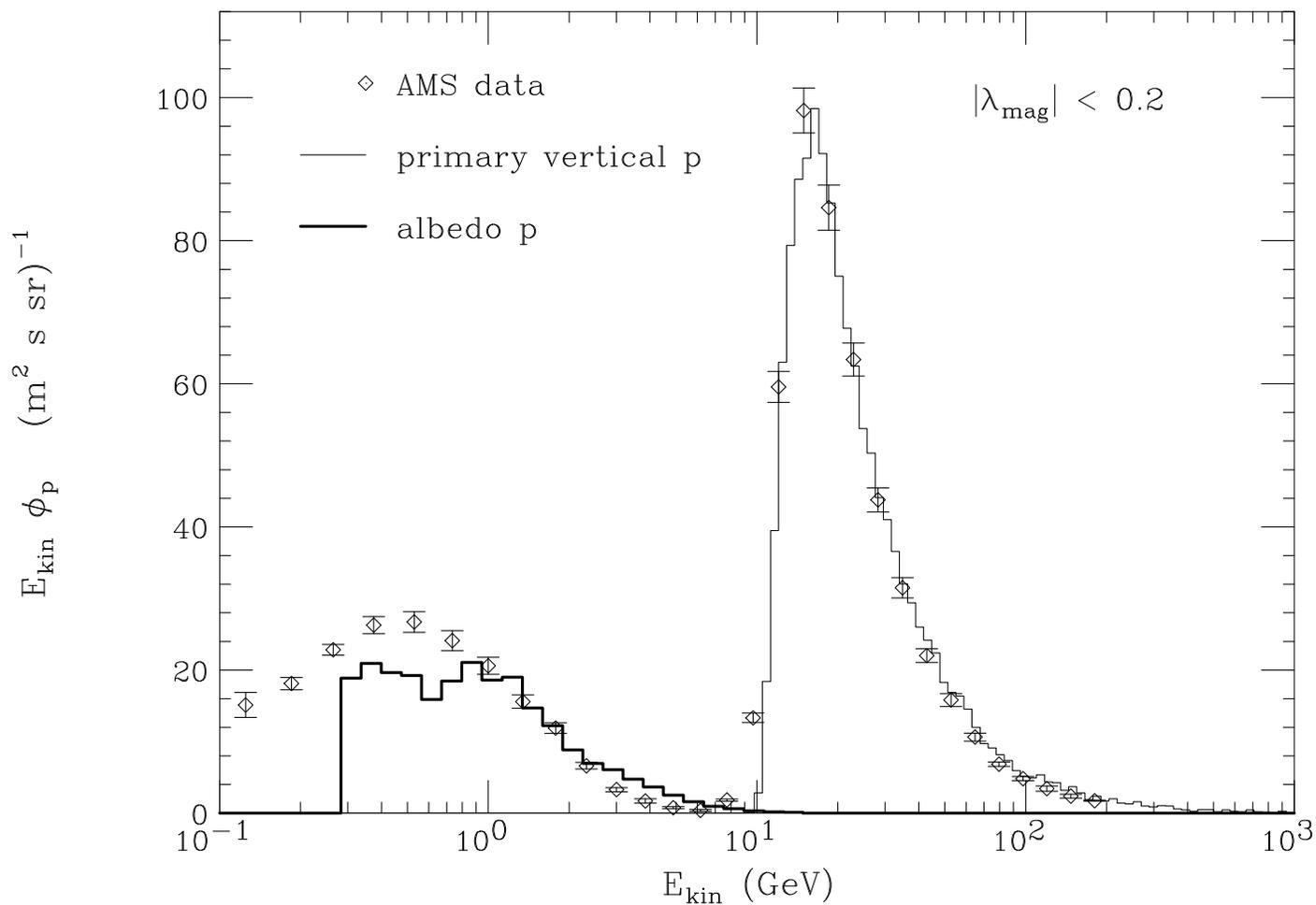,angle=90,height=12.9cm}}
\caption {\small  
The data points are the measurements of the vertical
proton flux by the AMS  detector \protect\cite{AMS-protons} in the
region of magnetic  longitude $|\lmag| < 0.2$.
The thin  histogram is a montecarlo calculation of the
primary  vertical proton
flux (averaged on 32$^\circ$  cone)   in the same  region.
The thick   histogram is a montecarlo estimate of
secondary protons  reaching the altitude of  380~Km   (always in the
same   region of magnetic latitude).
The MC calculation was performed only for $E_{\rm kin} > 0.290$~GeV
\label{fig:flux}  }
\end{figure}

\begin{figure} [t]
\centerline{\psfig{figure=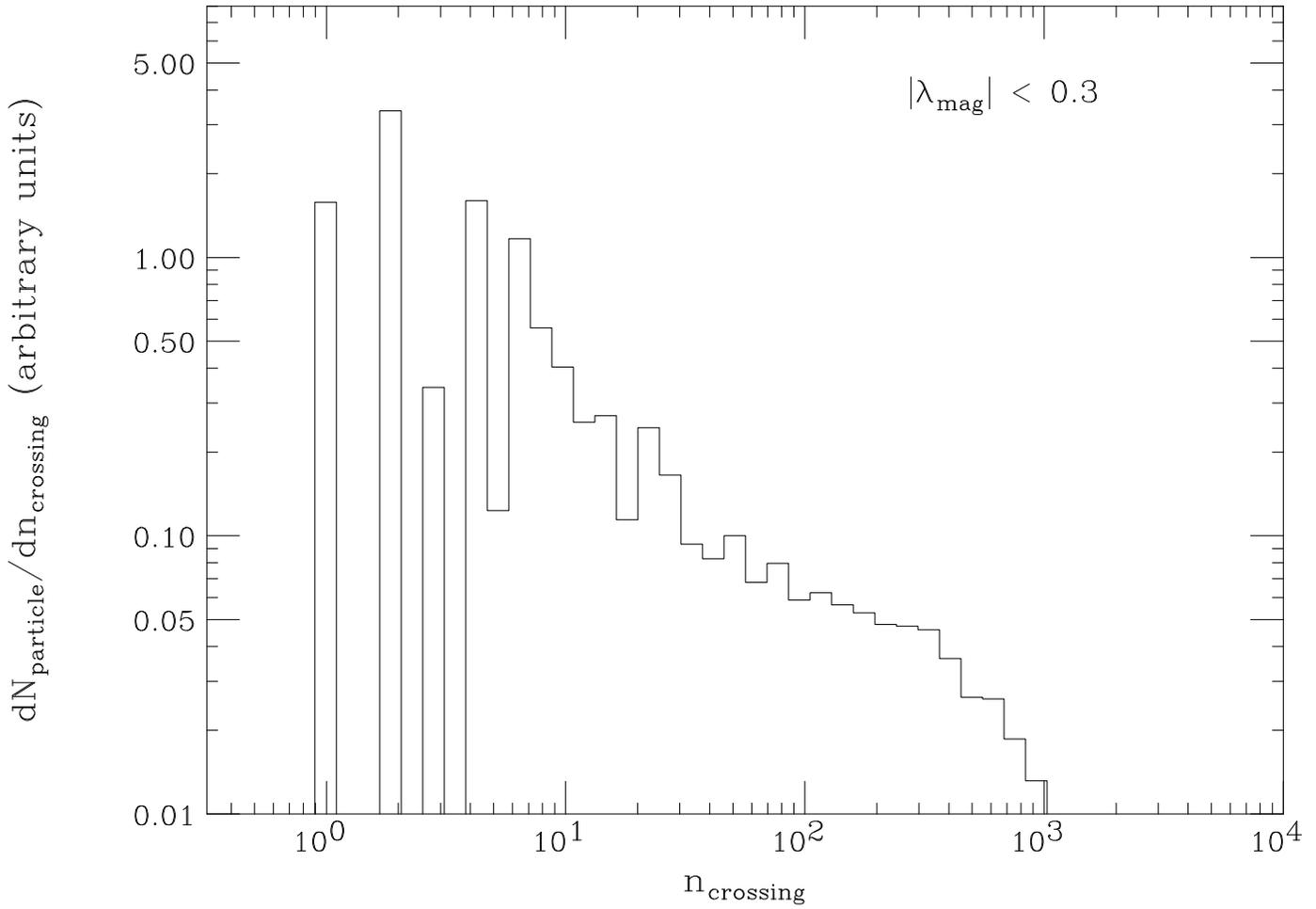,angle=90,height=12.9cm}}
\caption {\small  
Dis
tribution  of the number of crossings
that  albedo particle have with a surface 
of constant altitude  ($h = 380$~Km)  and magnetic  latitude
$|\lmag| \le 0.3$.
\label{fig:n0}  }
\end{figure}

\begin{figure} [t]
\centerline{\psfig{figure=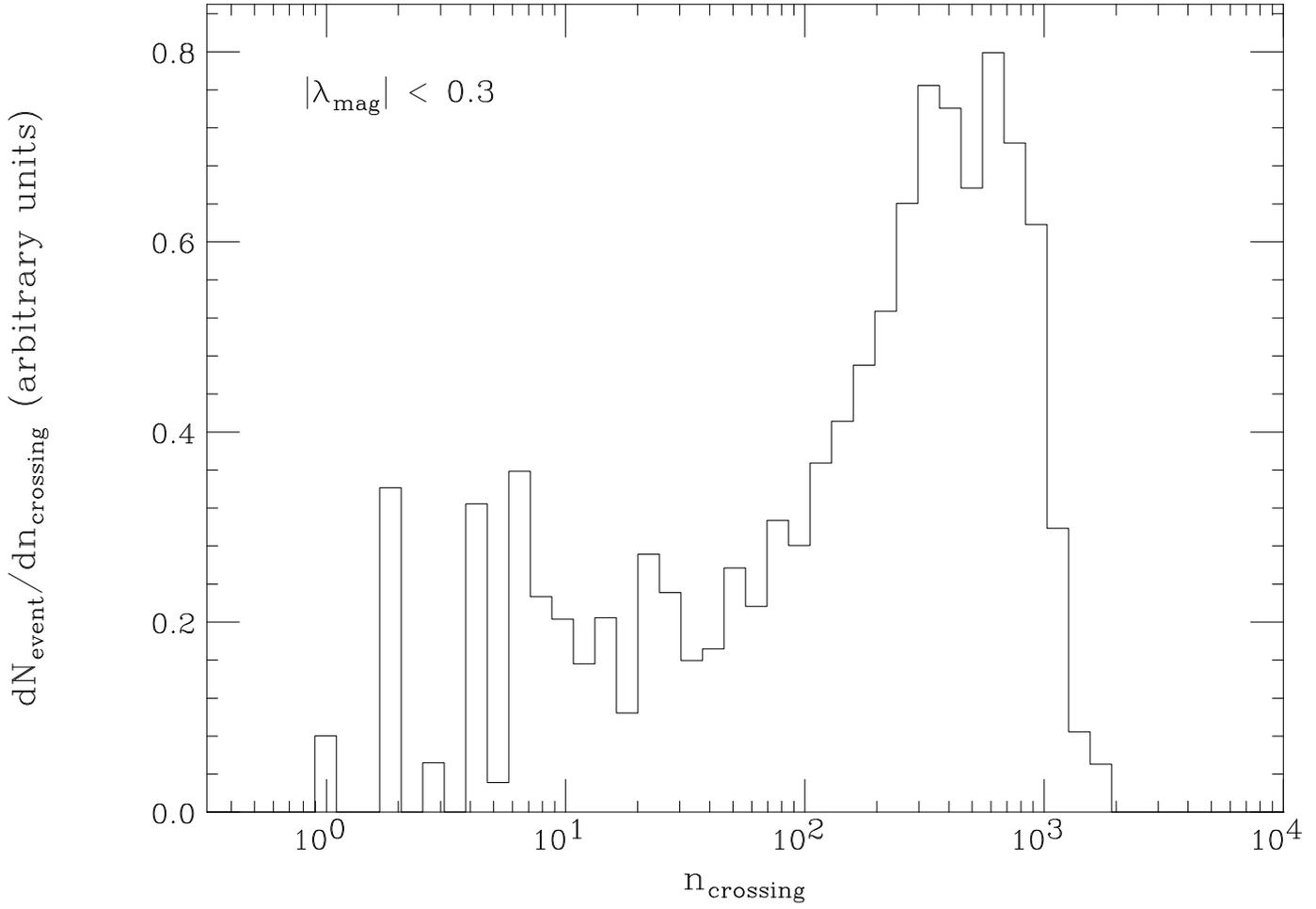,angle=90,height=12.9cm}}
\caption {\small  
Relative contribution of particles  with $N_{\rm crossings}$ to
the $p$ flux
in the magnetic  equatorial region 
($|\lmag| \le 11^\circ$)  at  an altitude  $h = 390$~Km.
\label{fig:n1}  }
\end{figure}

\begin{figure} [t]
\centerline{\psfig{figure=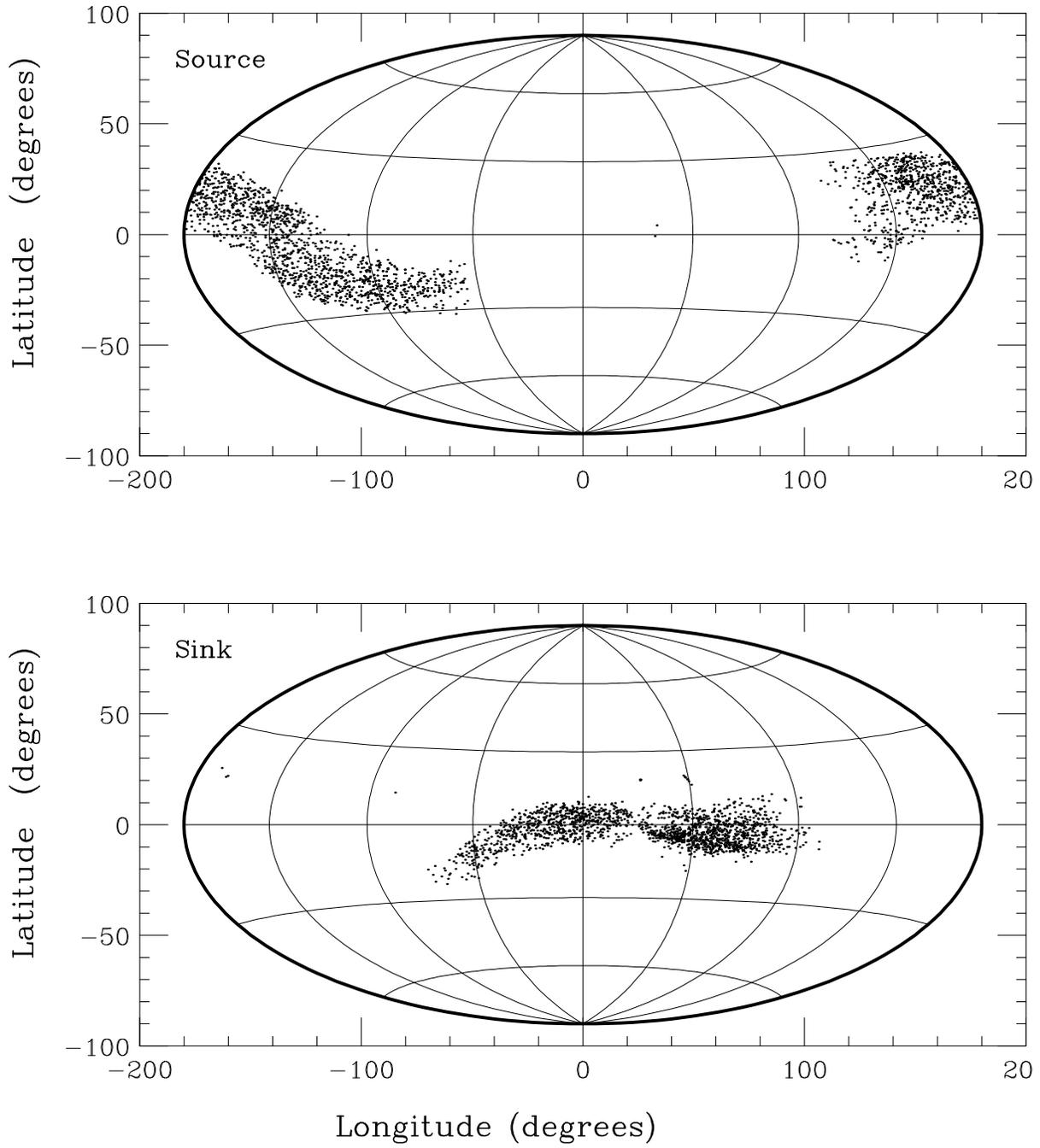,angle=90,height=18.0cm}}
\caption {\small  
In the top panel  the dots  represent  the points of origin of
long--lived  albedo protons.
In the bottom  panel  the dots  represent  the 
absorption  points  of 
long--lived  albedo protons.
\label{fig:points}  }
\end{figure}

\begin{figure} [t]
\centerline{\psfig{figure=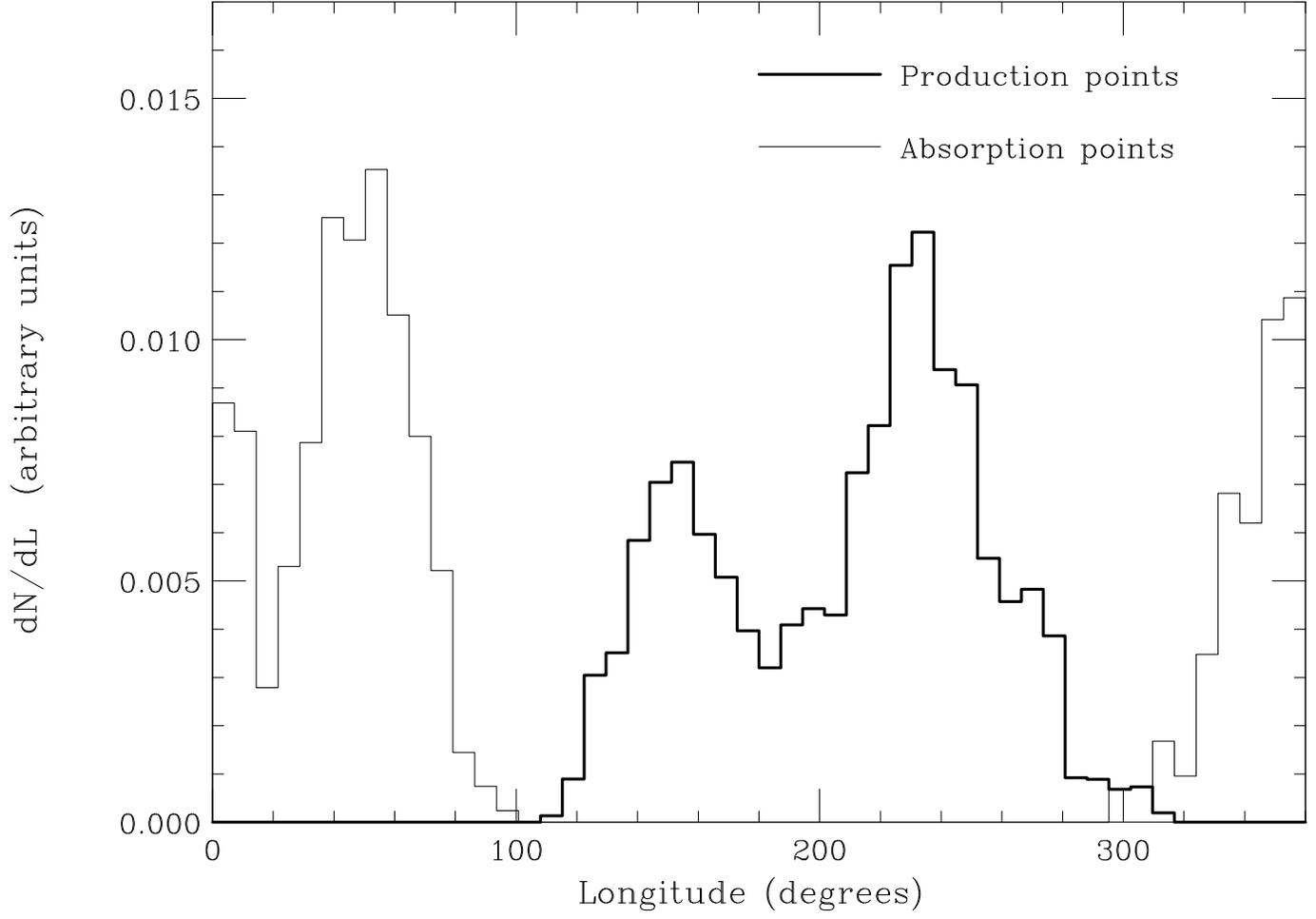,angle=90,height=12.9cm}}
\caption {\small  
Histogram  of the longitude  of  the production points 
of  long--lived albedo  protons
detected in the magnetic  equatorial region 
($|\lmag| \le 11^\circ$).
\label{fig:longitude}  }
\end{figure}

\begin{figure} [t]
\centerline{\psfig{figure=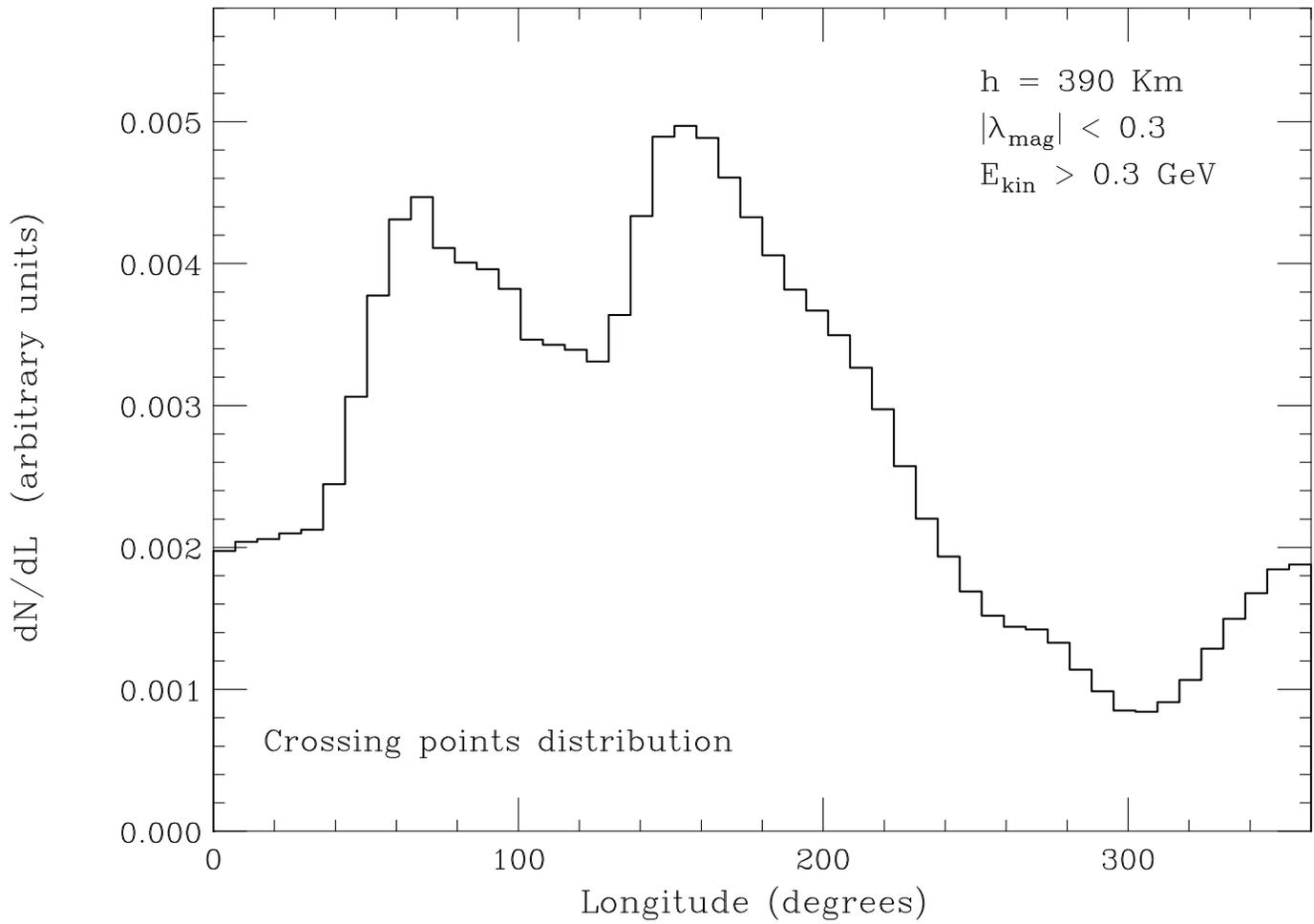,angle=90,height=12.9cm}}
\caption {\small  Longitude dependence of the $p$ albedo  flux
in the region  $|\lmag| < 0.2$   for 
$h = 380$~Km  and $E_k > 0.3$~GeV. 
\label{fig:long1}  }
\end{figure}

\begin{figure} [t]
\centerline{\psfig{figure=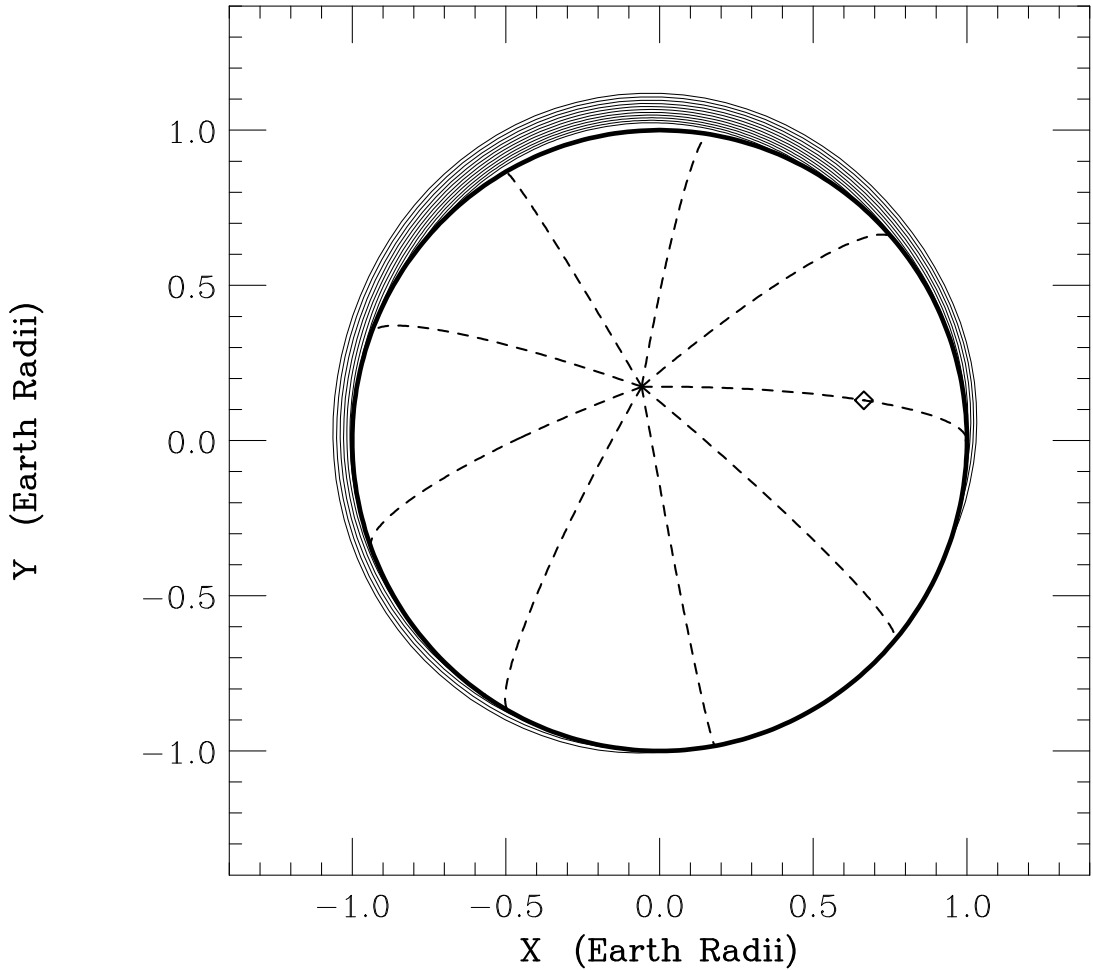,angle=90,height=13.0cm}}
\caption {\small
The figure shows the magnetic equatorial  plane    of  the Earth.
The dipole axis  is defined according to the first three terms of the  IGRF
expansion.
The thick  circle shows the surface of the Earth, 
the dashed  lines  are meridian lines, and the diamond  indicates
the  position of the Greenwich  observatory.
The  thin  solid lines 
are line of  constant  field.
It can be seen  that the lines   in first  approximation  are circles
centered  not on the Earth's  center, but on a point 
with $r/R_\oplus  \sim   0.06$  and   azimuth  $\phi \sim 120^\circ$.
\label{fig:equator2}  }
\end{figure}

\begin{figure} [t]
\centerline{\psfig{figure=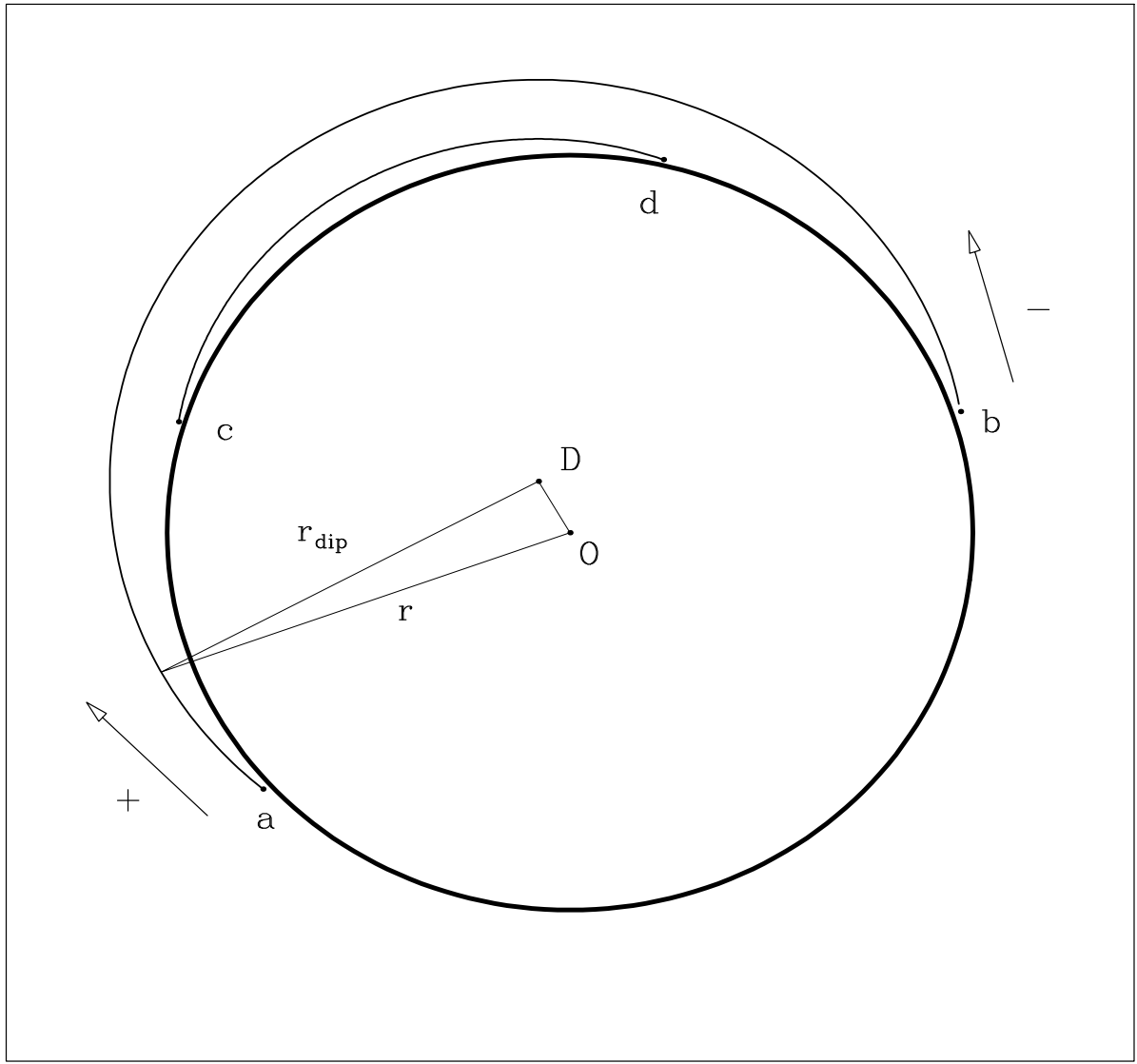,angle=90,height=15cm}}
\caption {\small  Illustration  of the properties of
 the    particle  trajectories confined in the
equatorial plane of an offset  dipole model.
The thick  circle represents the intersection of the Earth's  surface
with the dipole    equatorial plane.  Both the Earth's  center
(point $O$) and the dipole center  (point $D$)  are  on this plane.
The  thin  circles   represent the    trajectories of the guiding
center   of the  charged  particles  orbits. 
These  trajectories  remain at a constant  distance $r_{\rm dip}$ from 
the dipole center, and therefore  have  a variable  distance  $r$ from
the Earth's  center. 
Positively charged  particles  produced  at  the points  $a$ and $c$   will
drift   westward   and  be reabsorbed  at   points $b$ and $d$.
Negatively  charged  particles    drift in the opposite
(eastward)  direction, therefore  if produced  
at  the points  $a$ and $c$
will  be rapidly  reabsorbed,   while    if produced  at the points
$b$  or $d$  can   have    a long flight  time  
reaching points  $a$  and $b$.
\label{fig:eccentric}  }
\end{figure}

\begin{figure} [t]
\centerline{\psfig{figure=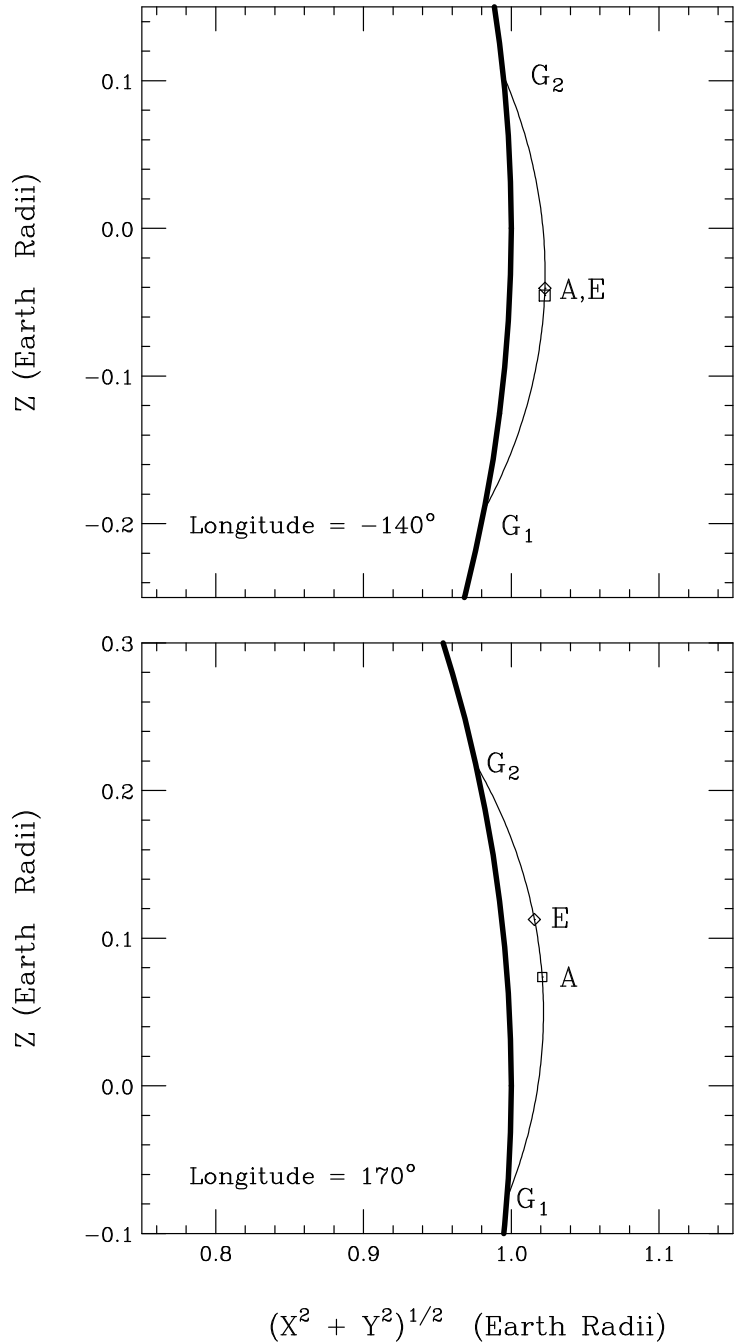,angle=90,height=18cm}}
\caption {\small  
Drawing of  two field  lines  of the geomagnetic  field.
The point labeled with  $A$
 in the top (bottom panel)   has  altitude  $h = 150$~Km and
longitude  of $-140^\circ$  (170$^\circ$).
In both cases  they are  the points  with  highest  $h$  along their
field line.
The points  labeled  as $E$ are the points on each line where 
 the magnetic  field 
has its  minimum value.
The field  line  intersect the Earth's surface 
at the points labeled   $G_1$ and $G_2$.
\label{fig:mirror}  }
\end{figure}

\begin{figure} [t]
\centerline{\psfig{figure=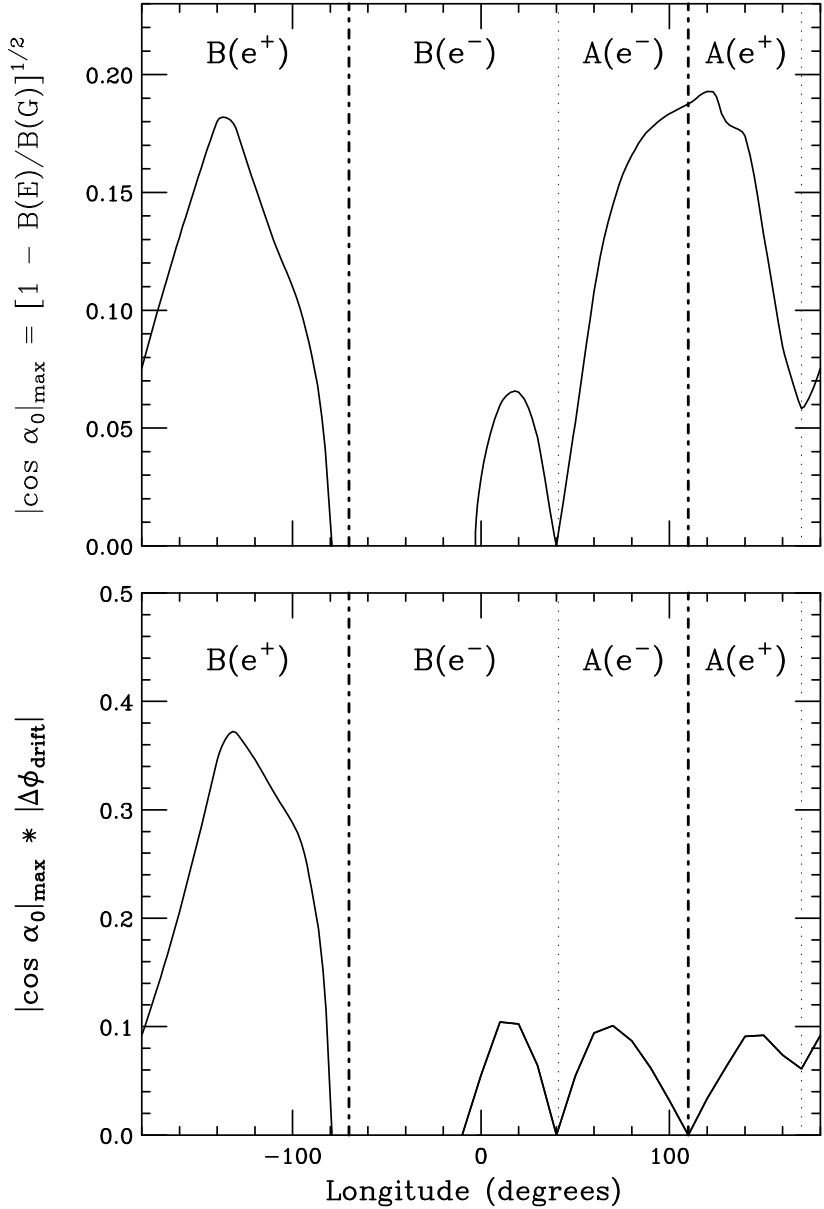,angle=90,height=16.2cm}}
\caption {\small  
The top  panel  shows 
the quantity $ \sqrt{1 - B_{\rm min}/B(G)}$
for all field  lines that have  maximum altitude  $h = 50$~Km.
Where  $B_{\rm min}$ is  the minimum  value of
the field  along the line, and $B(G)$ is the minimum  between
$B(G_1)$ and  $B(G_2)$ the   field  values at the two intersection points 
of a field line with the Earth's surface.
This  corresponds to the maximum  possible value   of $|\cos\alpha_0|$
for  particles  that can oscillate  along the line  without hitting 
the Earth's surface.  The angle $\alpha_0$ is the pitch angle of the
particle  at the minimum field point $E$.
The  field  lines  are identified  by the  longitude
of the equator point.
The vertical  dot--dashed  lines  separate the longitude  regions
for the injection  of positive and  negative particles.
The  vertical dotted  lines    separate   regions labeled  $A$ and $B$ in the
AMS  analysis \protect \cite{AMS-leptons}.
The bottom panel  shows  the product 
$|\cos \alpha_0|_{\rm max} \times |\Delta \varphi_{\rm drift}|$
where  $|\Delta \varphi_{\rm drift}|$  is the  longitude  drift
of a charged particle  injected into a  ``long''  trajectory
from a point with  longitude $\varphi$ (equation (\ref{eq:drift-interval})).
\label{fig:mirror1}  }
\end{figure}

\begin{figure}
\epsfig{figure=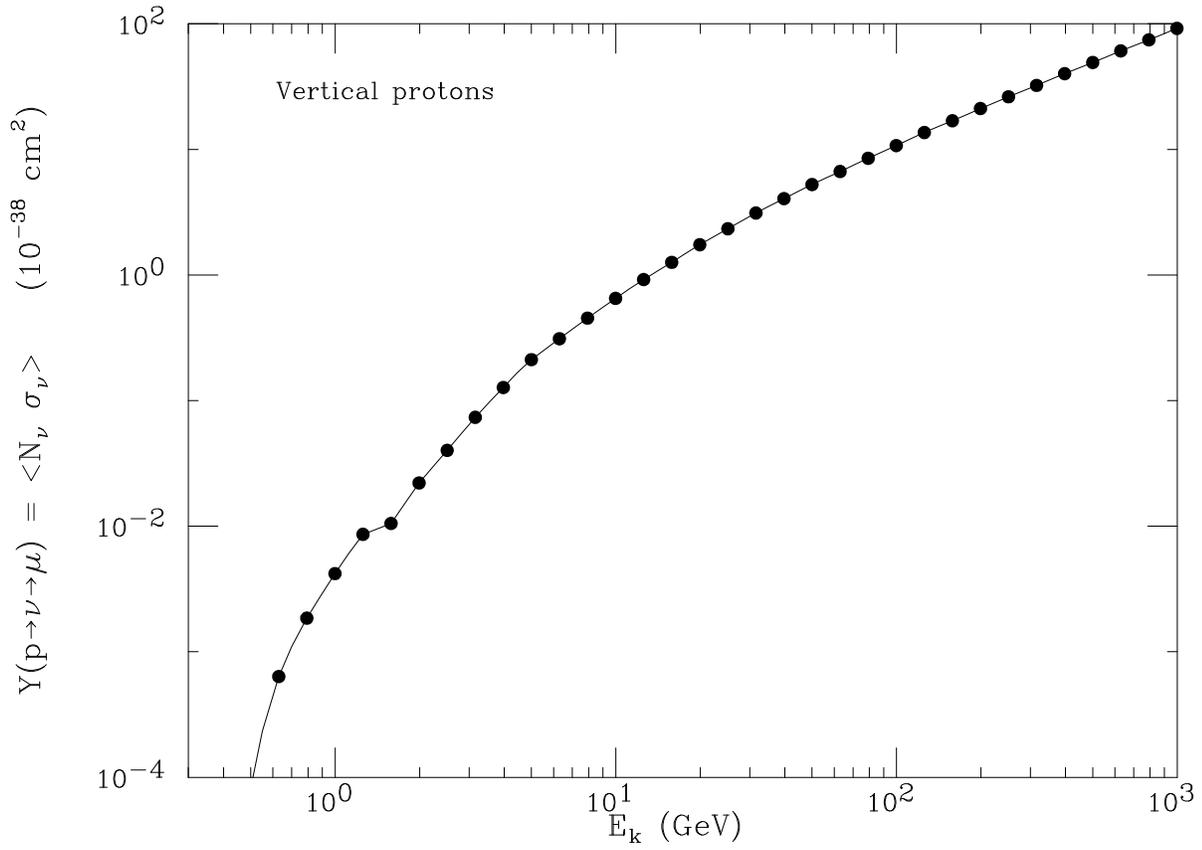,angle=90,height=11.3cm}
\caption {\small
The points  show the calculated  yield 
of $\nu$--induced  $\mu$--like events
for vertical  primary protons.
The yield  was calculated  using  Bartol  shower  model
 \protect\cite{Bartol}  and integrating  over all
$\nu$ production directions.
\label{fig:nu_yield} }
\end{figure}

\begin{figure}
\epsfig{figure=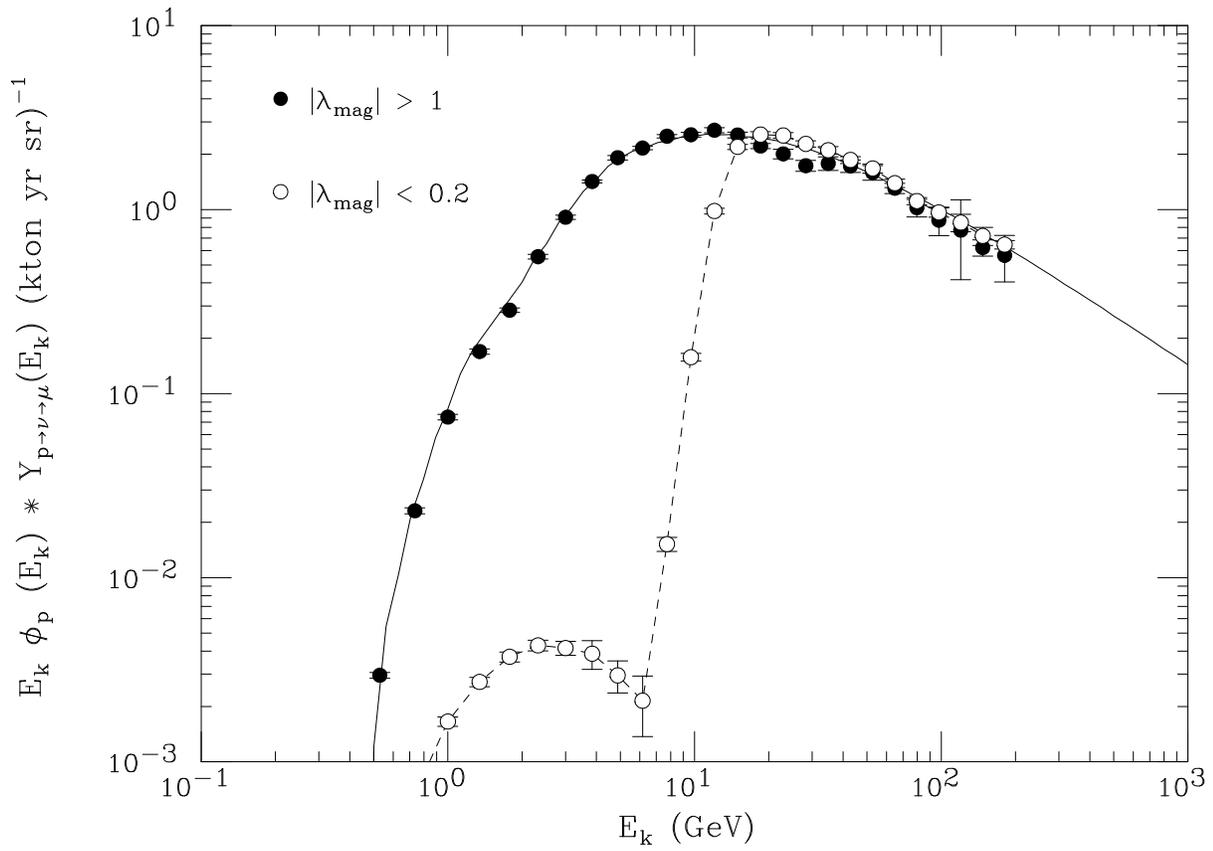,angle=90,height=11.3cm}
\caption {\small
Convolution of the   neutrino  event  yield
for  vertical protons with  the fluxes  observed
at high altitude  by the AMS  detector in  two
regions  of   geomagnetic latitude.
\label{fig:response} }
\end{figure}

For  illustration in  fig.~\ref{fig:traj1},\ref{fig:traj2}
 and~\ref{fig:ex1} 
we  show   examples 
of  trajectories  of   charged  particles    in the Earth's  field.
These particles  are examples   of ``long lived''
particles  generated  as  secondary protons in  the interaction 
of cosmic ray protons  in the atmosphere,  and have a
starting point  in a point   close to the Earth'surface where the 
density is large (at 
at an altitude of 40--50~Km  above sea level).
The  trajectories have been  calculated integrating   numerically
the  classical  equations of motion:
$d\vec{p}/dt = e \vec{\beta} \wedge \vec{B}$
in the IGRF  field.   In each figure two panels   show
a  \{$X$,$Y$\} (or equatorial)   projection (that  illustrate the
longitude  drift) and a  \{$\sqrt{X^2 + Y^2}$,$Z$\}  projection
(that illustrates  the   latitude  bouncing).
The  coordinates  are standard
 geographical  coordinates.

The  first  example (fig.~\ref{fig:traj1})) 
is  a particle    with
momentum 2.31~GeV
created  at  a latitude   $\lambda \simeq
-53^\circ$ in the southern hemisphere.
In the  \{$\sqrt{X^2 + Y^2}$,$Z$\}  projection it can be  clearly
that the particle  spirals  following the field   lines, 
performing   two ``bounces''.  The particle  cannot 
perform  one  additional  bounce  because  the bouncing point
is   ``inside the Earth'', and  the particle is  absorbed.
During the  flight the particle  drifts  uniformly westard,
as it is shown in the upper panel.

The trajectory  of the particle in the example 
of fig.~\ref{fig:traj2}
was    studied   for  a pathlength of
$10^{6}$~Km, before  interrupting the 
integration. 
In this  case  the particle  performs  
 many ``bounces''  and  drifts  in longitude for more than $2\pi$.

The  third  example in fig.~\ref{fig:ex1},
  shows the  trajectory  of  a particle
that remains close to the  equatorial plane of the field.
It  can  be noted  that the guiding center of the trajectory, travels
in an approximately  circular  motion  with a  center that 
does  not coincide with the Earth's  one.
This  is  a consequence  of the ``offset'' of the dipole
component   of the geomagnetic  field.

\subsection{Magnetic  shells}
\label{subsec:shells}
The most powerful mathematical  instrument to
describe the  motion of trapped  particles in the Earth  magnetic  field 
is the  concept of   ``magnetic  shells''.
This concept can be easily  illustrated 
 in the case  of a dipole field.
As  discussed  above, the motion  of  charged  particles   trapped in 
a  dipole  field can be  regarded  as the superposition of 
 a circular motion in a  plane  perpendicular  to the
local magnetic field,  around  a ``guiding center''
that  has a  much slower motion.
The``guiding center'  motion can be
analysed as an  oscillation   along a   guiding line
that  corresponds to a field  line,
and  a still  much  slower  rotation of the guiding line
  around the polar axis.
The  motion of the guiding  center  defines  therefore 
a surface that  can be called
a ``magnetic  shell''.  Each one of these shells 
corresponds to    the surface  generated by the rotation around the dipole 
axis of  a field line. Explicitely,  the  magnetic shells
have the  form $r = r_0 \, \cos^2 \lambda$, and can be labeled 
with  the parameter $r_0$.
The set of the mirror  points $M$ and $M^*$  
(in the north and south hemispheres) where each  individual  particle
``bounces'', that is  inverts the direction of the motion along the
field  line, have  a constant value of the magnetic  field $B$.

In the real   geomagnetic  field  
the motion of  trapped  charged  
has   qualitatively  very much the
same  structure as in the  dipole  case. To a good approximation
this  motion  can  again 
be  regarded   as  the superposition  of  a gyration and  a
motion  of the guiding  center.
This last motion  can again  be  analysed
 as an  oscillation 
along a field  line  between mirror  points  that have  a constant  value 
of  the  magnetic  field, and  a slower drift in longitude.
It can be demonstrated  (see \cite{Rossi-Olbert}) that  if
  a  charged  particle starts  oscillating along  a particular
field  line, after  drifting in  longitude through 360$^\circ$
it will  return to the field  line  from  which  it started.
Therefore
the  set of field lines   along which   a particle  oscillate
defines  again  a surface that  closes  upon itself as in the case of the 
dipole field\footnote{
This   decomposition  of the motion  of trapped  charged  particles
in three quasi--periodic
components  is connected   to the  existence of  three
adiabatic   invariants   for the  hamiltonian  of the  system
(see ref.\cite{Rossi-Olbert}  for a discussion).}.
 This  surface  can be again   considered  as a
``magnetic  shell'' and  is  called  an  ``$L$--shell''.
The  trajectory of the  guiding  center 
of particles  that have  latitude oscillations  with
different amplitudes,
 span  different    bands  around the ``equator''  of the shell.
In the case of  a  centered dipole  field  the  $L$  parameter  corresponds  to
the value  $r_0/R_\oplus$; 
for  an exact definition of the $L$ parameter 
 in the general  case see  for example  reference \cite{Rossi-Olbert}.
The  magnetic  field  lines that belong to the same
shell  to a  good approximation have the property that the  minimum value
of the field  along the line is  constant.

Disregarding the difference  the  positions of the particle  and
of its  guiding  center, the   motion of a trapped charged  particle 
are  therefore  confined  to   a  well defined   surface, that is  
the part  of an appropriate  magnetic  shell,  where  the magnetic  field
is  lower than  a maximum value $B^*$  with the   set of  points  on the 
shell where the  field has the value  $B^*$  corresponding   to the  set of
mirror points  for the particle trajectory.
The set of all possible trajectories can  be  classified according to 
the two parameters  ($L$ and $B^*$)  that  defines  the shell, 
and the set of mirror  points on the shell.

The  magnetic shells  
of  the real  geomagnetic    field   are not  exactly symmetric 
with respect to the  Earth center. Qualitatively the most important
feature is  an offset, such  that the equator of 
the most internal shells 
intersect the surface of the  Earth.
These  intersection  regions  are both the source and  sink regions
of the long lived albedo particle.
Since positively and negatively  charged  particles  drift in opposite
directions  the sources  of positive particles are the sink
 of negative particles  and viceversa.

\clearpage

\clearpage


\


\


\end{document}